\documentclass[aps,prd,preprint,tightenlines,superscriptaddress,byrevtex]{revtex4}
\usepackage{graphicx} 
\usepackage{amsmath}

\graphicspath{{ps}}

\begin{document}


\preprint{\vbox{ \hbox{   }
                 \hbox{BELLE-CONF-0845}
}}

\title{ \quad\\[0.5cm] \boldmath Measurement of the form factors of
  the decay $B^0\to D^{*-}\ell^+\nu_\ell$ and determination of the CKM
  matrix element $|V_{cb}|$}

\affiliation{Budker Institute of Nuclear Physics, Novosibirsk}
\affiliation{Chiba University, Chiba}
\affiliation{University of Cincinnati, Cincinnati, Ohio 45221}
\affiliation{Department of Physics, Fu Jen Catholic University, Taipei}
\affiliation{Justus-Liebig-Universit\"at Gie\ss{}en, Gie\ss{}en}
\affiliation{The Graduate University for Advanced Studies, Hayama}
\affiliation{Gyeongsang National University, Chinju}
\affiliation{Hanyang University, Seoul}
\affiliation{University of Hawaii, Honolulu, Hawaii 96822}
\affiliation{High Energy Accelerator Research Organization (KEK), Tsukuba}
\affiliation{Hiroshima Institute of Technology, Hiroshima}
\affiliation{University of Illinois at Urbana-Champaign, Urbana, Illinois 61801}
\affiliation{Institute of High Energy Physics, Chinese Academy of Sciences, Beijing}
\affiliation{Institute of High Energy Physics, Vienna}
\affiliation{Institute of High Energy Physics, Protvino}
\affiliation{Institute for Theoretical and Experimental Physics, Moscow}
\affiliation{J. Stefan Institute, Ljubljana}
\affiliation{Kanagawa University, Yokohama}
\affiliation{Korea University, Seoul}
\affiliation{Kyoto University, Kyoto}
\affiliation{Kyungpook National University, Taegu}
\affiliation{\'Ecole Polytechnique F\'ed\'erale de Lausanne (EPFL), Lausanne}
\affiliation{Faculty of Mathematics and Physics, University of Ljubljana, Ljubljana}
\affiliation{University of Maribor, Maribor}
\affiliation{University of Melbourne, School of Physics, Victoria 3010}
\affiliation{Nagoya University, Nagoya}
\affiliation{Nara Women's University, Nara}
\affiliation{National Central University, Chung-li}
\affiliation{National United University, Miao Li}
\affiliation{Department of Physics, National Taiwan University, Taipei}
\affiliation{H. Niewodniczanski Institute of Nuclear Physics, Krakow}
\affiliation{Nippon Dental University, Niigata}
\affiliation{Niigata University, Niigata}
\affiliation{University of Nova Gorica, Nova Gorica}
\affiliation{Osaka City University, Osaka}
\affiliation{Osaka University, Osaka}
\affiliation{Panjab University, Chandigarh}
\affiliation{Peking University, Beijing}
\affiliation{Princeton University, Princeton, New Jersey 08544}
\affiliation{RIKEN BNL Research Center, Upton, New York 11973}
\affiliation{Saga University, Saga}
\affiliation{University of Science and Technology of China, Hefei}
\affiliation{Seoul National University, Seoul}
\affiliation{Shinshu University, Nagano}
\affiliation{Sungkyunkwan University, Suwon}
\affiliation{University of Sydney, Sydney, New South Wales}
\affiliation{Tata Institute of Fundamental Research, Mumbai}
\affiliation{Toho University, Funabashi}
\affiliation{Tohoku Gakuin University, Tagajo}
\affiliation{Tohoku University, Sendai}
\affiliation{Department of Physics, University of Tokyo, Tokyo}
\affiliation{Tokyo Institute of Technology, Tokyo}
\affiliation{Tokyo Metropolitan University, Tokyo}
\affiliation{Tokyo University of Agriculture and Technology, Tokyo}
\affiliation{Toyama National College of Maritime Technology, Toyama}
\affiliation{Virginia Polytechnic Institute and State University, Blacksburg, Virginia 24061}
\affiliation{Yonsei University, Seoul}
  \author{I.~Adachi}\affiliation{High Energy Accelerator Research Organization (KEK), Tsukuba} 
  \author{H.~Aihara}\affiliation{Department of Physics, University of Tokyo, Tokyo} 
  \author{D.~Anipko}\affiliation{Budker Institute of Nuclear Physics, Novosibirsk} 
  \author{K.~Arinstein}\affiliation{Budker Institute of Nuclear Physics, Novosibirsk} 
  \author{T.~Aso}\affiliation{Toyama National College of Maritime Technology, Toyama} 
  \author{V.~Aulchenko}\affiliation{Budker Institute of Nuclear Physics, Novosibirsk} 
  \author{T.~Aushev}\affiliation{\'Ecole Polytechnique F\'ed\'erale de Lausanne (EPFL), Lausanne}\affiliation{Institute for Theoretical and Experimental Physics, Moscow} 
  \author{T.~Aziz}\affiliation{Tata Institute of Fundamental Research, Mumbai} 
  \author{S.~Bahinipati}\affiliation{University of Cincinnati, Cincinnati, Ohio 45221} 
  \author{A.~M.~Bakich}\affiliation{University of Sydney, Sydney, New South Wales} 
% ...
\collaboration{The Belle Collaboration}

\begin{abstract}
  This paper describes a determination of the Cabibbo-Kobayashi-Maskawa
matrix element~$|V_{cb}|$ using the decay $B^0\to
D^{*-}\ell^+\nu_\ell$. We perform a combined measurement of this
quantity and of the form factors $\rho^2$, $R_1(1)$, and $R_2(1)$
which fully characterize this decay in the framework of heavy-quark
effective theory, based on 140 fb$^{-1}$ of Belle data collected near
the $\Upsilon(4S)$~resonance. The results, based on about 69,000
reconstructed $B^0\to D^{*-}\ell^+\nu_\ell$ decays, are
$\rho^2=1.293\pm 0.045\pm 0.029$, $R_1(1)=1.495\pm 0.050\pm 0.062$,
$R_2(1)=0.844\pm 0.034\pm 0.019$ and $\mathcal{F}(1)|V_{cb}|=34.4\pm
0.2\pm 1.0$. The $B^0\to D^{*-}\ell^+\nu_\ell$~branching fraction is
found to be $(4.42\pm 0.03\pm 0.25)\%$. For all these numbers, the
first error is the statistical and the second is the systematic
uncertainty. All results are preliminary.

\end{abstract}


\maketitle

\tighten

{\renewcommand{\thefootnote}{\fnsymbol{footnote}}}
\setcounter{footnote}{0}

\section{Introduction}

The study of the decay~$B^0\to D^{*-}\ell^+\nu_\ell$ is an important
item on the $B$~physics agenda for many reasons. First, the total rate
is proportional to the magnitude of the Cabibbo-Kobayashi-Maskawa
(CKM) matrix element $V_{cb}$~\cite{Kobayashi:1973fv} squared. We can
thus determine this quantity from the measurement of this
decay. Second, $B^0\to D^{*-}\ell^+\nu_\ell$ is a major background for
charmless semileptonic $B$~decays or semileptonic $B$~decays with
large missing energy. A precise knowledge of the form factors of this
decay will thus help reducing systematic uncertainties in these
analyses.

Our analysis technique follows closely previous studies of this decay
using
$e^+e^-\to\Upsilon(4S)$~data~\cite{Briere:2002ew,Abe:2001cs,Aubert:2007rs},
{\it i.e.}, we reconstruct the $D^{*}$~meson and the charged lepton
only, without making any requirement on the other $B$~meson in the
event. The main difference to earlier
analyses~\cite{Briere:2002ew,Abe:2001cs} is that we measure the CKM
matrix element and all three HQET form factors of this decay
simultaneously. Also, by using a novel reconstruction technique of the
$B$~meson 4-momentum, we achieve a better resolution in the kinematic
variables describing the $B^0\to D^{*-}\ell^+\nu_\ell$~decay, which
translates into an improved determination of the HQET form factors.

\section{Theoretical framework}

\subsection{Kinematic variables} \label{sec:2a}

The decay $B^0\to D^{*-}\ell^+\nu_\ell$~\cite{ref:0} proceeds chiefly
through the tree-level transition shown in Fig.~\ref{fig:1}. Its
kinematics can be fully characterized by four variables:
\begin{figure}
  \begin{center}
    \includegraphics[width=0.4\columnwidth]{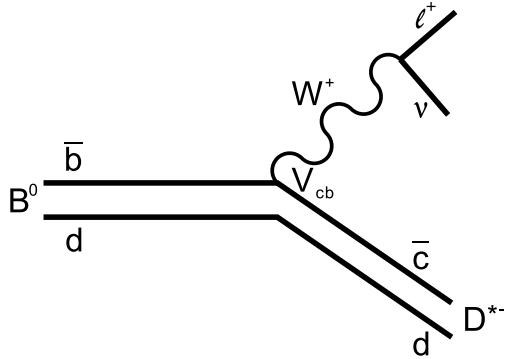}
  \end{center}
  \caption{Quark-level Feynman diagram for the decay $B^0\to
    D^{*-}\ell^+\nu_\ell$.} \label{fig:1}
\end{figure}

The first one is $w$, defined by
\begin{equation}
  w=\frac{p_B\cdot p_{D^*}}{m_B
    m_{D^*}}=\frac{m_B^2+m_{D^*}^2-q^2}{2m_Bm_{D^*}}~,
\end{equation}
where $m_B$ and $m_{D^*}$ are the masses of the $B$ and the $D^*$
mesons (5.2794~GeV and 2.010~GeV, respectively~\cite{Yao:2006px}), $p_B$
and $p_{D^*}$ are their four-momenta, and $q^2=(p_\ell+p_\nu)^2$. In
the $B$~rest frame, approximately equal to the
$\Upsilon(4S)$~center-of-mass (c.m.) frame, the expression for $w$
reduces to the Lorentz boost $\gamma_{D^*}=E_{D^*}/m_{D^*}$. The
ranges of $w$ and $q^2$ are restricted by the kinematics of the decay,
with $q^2 = 0$ corresponding to
\begin{equation}
  w_\mathrm{max}=\frac{m_B^2 + m_{D^*}^2}{2m_B m_{D^*}} \approx 1.504~,
\end{equation}
and $w_\mathrm{min}=1$ to
\begin{equation}
q_\mathrm{max}^2=(m_B-m_{D^*})^2 = 10.69~\mathrm{GeV}^2~.
\end{equation}
The point~$w=1$ is also refered to as zero recoil.

\begin{figure}
  \begin{center}
    \includegraphics[width=0.8\columnwidth]{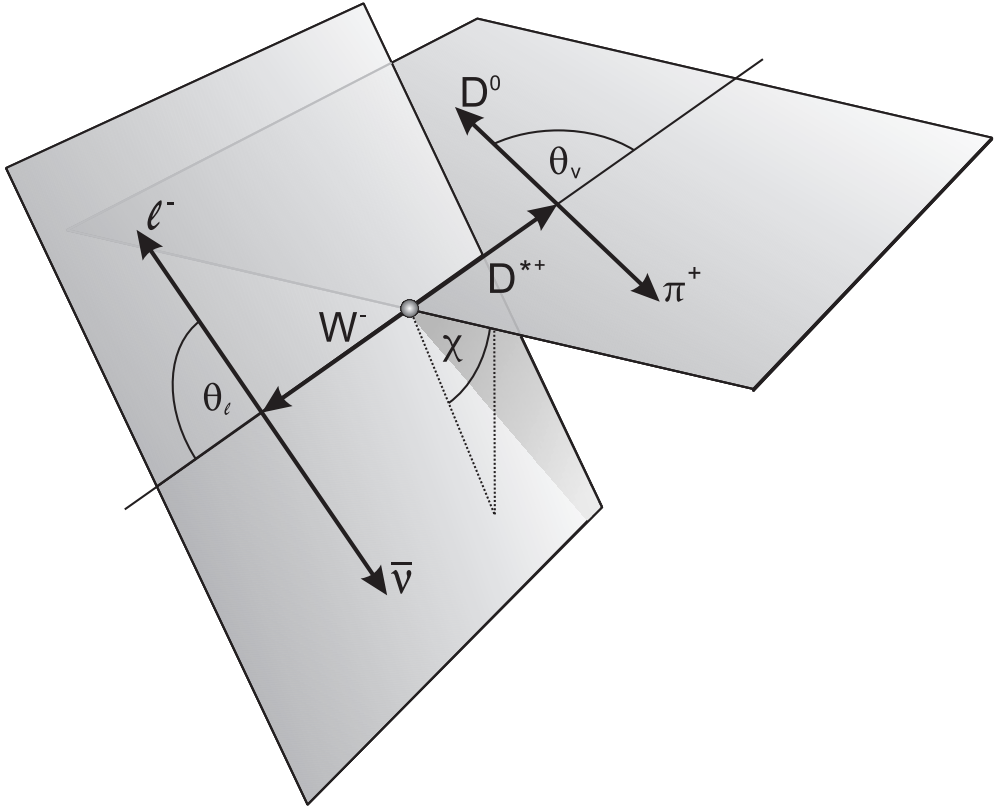}
  \end{center}
  \caption{Definition of the angles $\theta_\ell$, $\theta_V$ and
    $\chi$ for the decay $B^0\to D^{*-}\ell^+\nu_\ell$, $D^{*-}\to\bar
    D^0\pi_s^-$.} \label{fig:2}
\end{figure}

The remaining three variables are the angles shown in Fig.~\ref{fig:2}:
\begin{itemize} 
\item{$\theta_\ell$, the angle between the direction of the lepton
  in the virtual $W$~rest frame and the direction of the $W$ in the
  $B$~rest frame;}
\item{$\theta_V$, the angle between the direction of the $D$~meson in the
  $D^*$ rest frame and the direction of the $D^*$~meson in the $B$
  rest frame;}
\item{$\chi$, the angle between the $D^*$ and $W$~decay planes in the
  $B$~rest frame.}
\end{itemize} 

\subsection{Four-dimensional decay distribution}

The Lorentz structure of the $B^0\to D^{*-}\ell^+\nu_\ell$~decay
amplitude can be expressed in terms of three helicity amplitudes
($H_{+}$, $H_{-}$, and $H_{0}$), which correspond to the three
polarization states of the $D^*$, two transverse and one
longitudinal. For low-mass leptons (electrons and muons), these
amplitudes are expressed in terms of the three functions $h_{A_1}(w)$,
$R_1(w)$, and $R_2(w)$~\cite{Neubert:1993mb}
\begin{equation}
  H_i(w)=m_B\frac{R^*(1-r^2)(w+1)}{2\sqrt{1-2wr+r^2}}h_{A_1}(w)\tilde{H}_i(w)~,
\end{equation}
where
\begin{eqnarray}
  \tilde{H}_{\mp} & = &
  \frac{\sqrt{1-2wr+r^2}\left(1\pm\sqrt{\frac{w-1}{w+1}}
    R_1(w)\right)}{1-r}~, \\
  \tilde{H}_0 & = & 1+\frac{(w-1)(1-R_2(w))}{1-r}~,
\end{eqnarray}
with $R^*=(2\sqrt{m_B m_{D^*}})/(m_B+m_{D^*})$ and
$r=m_{D^*}/m_B$. The functions $R_1(w)$ and $R_2(w)$ are defined in
terms of the axial and vector form factors as,
\begin{equation}
  A_2(w)=\frac{R_2(w)}{R^{*2}}\frac{2}{w+1}A_1(w)~,
\end{equation}
\begin{equation}
  V(w)=\frac{R_1(w)}{R^{*2}}\frac{2}{w+1}A_1(w)~.
\end{equation}
By convention, the function $h_{A_1}(w)$ is defined as
\begin{equation}
  h_{A_1}(w)=\frac{1}{R^*}\frac {2}{w+1} A_1(w)~.
\end{equation}
For $w\to 1$, the axial form factor $A_1(w)$ dominates, and in the
limit of infinite $b$- and $c$-quark masses, a single form factor
describes the decay, the so-called Isgur-Wise
function~\cite{Isgur:1989vq,Isgur:1989ed}.

The fully differential decay rate in terms of the three helicity
amplitudes is~\cite{Richman:1995wm}
\begin{equation}
  \begin{split}
    & \frac{\mathrm{d}^4\Gamma(B^0\to
      D^{*-}\ell^+\nu_\ell)}{\mathrm{d}w\mathrm{d}(\cos\theta_\ell)\mathrm{d}(\cos\theta_V)\mathrm{d}\chi}=\frac{6m_Bm_{D^*}^2}{8(4\pi)^4}\sqrt{w^2-1}(1-2wr+r^2)G_F^2|V_{cb}|^2\\
    & \times\big\{(1-\cos\theta_\ell)^2\sin^2\theta_VH^2_+(w)+(1+\cos\theta_\ell)^2\sin^2\theta_VH^2_-(w)\biggr. \\
    & +4\sin^2\theta_\ell\cos^2\theta_VH^2_0(w)-2\sin^2\theta_\ell\sin^2\theta_V\cos 2\chi H_+(w)H_-(w) \\
    & -4\sin\theta_\ell(1-\cos\theta_\ell)\sin\theta_V\cos\theta_V\cos\chi H_+(w)H_0(w) \\
    &
      +\biggl. 4\sin\theta_\ell(1+\cos\theta_\ell)\sin\theta_V\cos\theta_V\cos\chi H_-(w)H_0(w)\big\}~, \label{eq:2_1}
  \end{split}
\end{equation}
with $G_F=(1.16637\pm 0.00001)\times 10^{-5}$~GeV$^{-2}$. By
integrating this decay rate over all but one of the four variables,
$w$, $\cos\theta_\ell$, $\cos\theta_V$, or $\chi$, we obtain the four
one-dimensional decay distributions from which we will extract the
form factors. The differential decay rate as a function of $w$ is
\begin{equation}
  \frac{\mathrm{d}\Gamma}{\mathrm{d}w}=\frac{G^2_F}{48\pi^3}m^3_{D*}\big(m_B-m_{D^*}\big)^2\mathcal{G}(w)\mathcal{F}^2(w)|V_{cb}|^2~, \label{eq:2_2}
\end{equation}
where  
\begin{eqnarray*}
  \mathcal{F}^2(w)\mathcal{G}(w)=h_{A_1}^2(w)\sqrt{w-1}(w+1)^2\left\{2\left[\frac{1-2wr+r^2}{(1-r)^2}\right]\right.
  \\
  \left. \times\left[1+R_1(w)^2\frac{w-1}{w+1}\right]+\left[1+(1-R_2(w))\frac{w-1}{1-r}\right]^2\right\}~,
\end{eqnarray*}
and $\mathcal{G}(w)$ is a known phase space factor,
\begin{equation*}
  \mathcal{G}(w)=\sqrt{w^2-1}(w+1)^2\left[1+4\frac{w}{w+1}\frac{1-2wr+r^2}{(1-r)^2}\right].
\end{equation*}

In the infinite quark-mass limit, the heavy quark symmetry (HQS)
predicts $\mathcal{F}(1)=1$. Corrections to this limit have been
calculated in lattice QCD. A calculation, performed in the quenched
approximation, predicts (including a QED correction of 0.7\%)
$\mathcal{F}(1)=0.919^{+0.030}_{-0.035}$~\cite{Hashimoto:2001nb}. This
value is compatible with estimates based on non-lattice
methods~\cite{Uraltsev:2000qw}. A recent unquenched lattice result,
$\mathcal{F}(1)=0.930\pm 0.022$, is still
preliminary~\cite{Laiho:2007pn}.

\subsection{Form factor parameterization}

The heavy quark effective theory (HQET) allows to obtain a
parameterization of these form-factors. Perfect heavy quark symmetry
implies that $R_1(w)=R_2(w)=1$, {\it i.e.}, the form factors $A_2$ and
$V$ are identical for all values of $w$ and differ from $A_1$ only by
a simple kinematic factor. Corrections to this approximation have been
calculated in powers of $\Lambda_\mathrm{QCD}/m_b$ and the strong
coupling constant $\alpha_s$. Various parameterizations in powers of
$(w-1)$ have been proposed. Among the different predictions relating
the coefficients of the higher order terms to the linear term, we
adopt the following expressions derived by Caprini, Lellouch and
Neubert~\cite{Caprini:1997mu},
\begin{eqnarray}
  h_{A_1}(w) & = &
  h_{A_1}(1)\big[1-8\rho^2z+(53\rho^2-15)z^2-(231\rho^2-91)z^3\big]~,
  \label{eq:2_3} \\
  R_1(w) & = & R_1(1)-0.12(w-1)+0.05(w-1)^2~, \label{eq:2_4} \\ 
  R_2(w) & = & R_2(1)+0.11(w-1)-0.06(w-1)^{2}~, \label{eq:2_5}  
\end{eqnarray}
where $z=(\sqrt{w+1}-\sqrt{2})/(\sqrt{w+1}+\sqrt{2})$. The three
parameters $\rho^{2}$, $R_1(1)$, and $R_2(1)$, cannot be calculated;
they must be extracted from data.

\section{Experimental procedure}

\subsection{Data sample and event selection}

The data used in this analysis were taken with the Belle
detector~\cite{unknown:2000cg} at the KEKB asymmetric energy
$e^+e^-$~collider~\cite{Kurokawa:2001nw}. Belle is a large-solid-angle
magnetic spectrometer that consists of a three-layer silicon vertex
detector, a 50-layer central drift chamber (CDC), an array of
aerogel threshold Cherenkov counters (ACC), a barrel-like
arrangement of time-of-flight scintillation counters (TOF), and an
electromagnetic calorimeter (ECL) comprised of CsI(Tl) crystals
located inside a super-conducting solenoid coil that provides a 1.5~T
magnetic field. An iron flux-return located outside of the coil is
instrumented to detect $K_L^0$ mesons and to identify muons (KLM).

The data sample consists of 140~fb$^{-1}$ taken at the
$\Upsilon(4S)$~resonance, or $152\times 10^6$ $B\bar B$~events. Another
15~fb$^{-1}$ taken at 60~MeV below the resonance are
used to estimate the non-$B\bar B$ (continuum) background. The
off-resonance data is scaled by the integrated on- to off-resonance
luminosity ratio corrected for the $1/s$~dependence of the $q\bar
q$~cross-section.

Generic Monte Carlo samples equivalent to about three times the
integrated luminosity are used in this analysis. Monte Carlo simulated
events are generated with the {\tt evtgen} program~\cite{Lange:2001uf}
and full detector simulation based on {\tt GEANT}~\cite{Brun:1987ma}
is applied. QED bremsstrahlung in $B\to X\ell\nu$~decays is added
using the {\tt PHOTOS} package~\cite{Barberio:1993qi}.

Hadronic events are selected based on the charged track multiplicity
and the visible energy in the calorimeter. The selection is described
in detail elsewhere~\cite{Abe:2001hj}. We also apply a moderate cut on
the ratio of the second to the zeroth Fox-Wolfram
moment~\cite{Fox:1978vu}, $R_2 < 0.4$, to reject continuum events.

\subsection{Event reconstruction}

Charged tracks are required to originate from the interaction point by
applying the following selections on the impact parameters in $r\phi$
and $z$, $|dr|<2$~cm and $|dz|<4$~cm, respectively. Additionally, we
demand at least one associated hit in the SVD detector. For pion and
kaon candidates, the Cherenkov light yield from ACC, the
time-of-flight information from TOF and $dE/dx$ from CDC are required
to be consistent with the respective mass hypothesis.

Neutral $D$~meson candidates are searched for in the decays channels
$D^0\to K^-\pi^+$ and $D^0\to K^-\pi^+\pi^-\pi^+$. We fit the charged
tracks to a common vertex and reject the $D^0$~candidate if the
$\chi^2$-probability is below $10^{-3}$. The momenta of the charged
tracks are re-evaluated at the vertex and the $D^0$ 4-momentum is
calculated as their sum. The reconstructed $D^0$~mass is required to
lie within $\pm 3$ standard deviations from $m_{D^0}$, where one sigma
is about 4.5~MeV (4~MeV) for the one (three) pion mode.

The $D^0$~candidate is combined with a slow pion~$\pi^+_s$
(appropriately charged with respect to the kaon candidate) to form a
$D^{*+}$~candidate. No impact parameter and SVD hit requirements are
applied for $\pi_s$. Again, a vertex fit is performed and the same
vertex requirement is applied. The invariant mass difference between
the $D^*$ and the $D$~candidates, $\Delta m$, is required to lie within
144 and 147~MeV. Additional continuum suppression is achieved by
requiring a $D^*$~momentum below 2.45~GeV in the c.m.\ frame.

Finally, the $D^*$~candidate is combined with an oppositely charged
lepton (electron or muon). Electron candidates are identified using
the ratio of the energy detected in the ECL to the track momentum, the
ECL shower shape, position matching between track and ECL cluster, the
energy loss in the CDC and the response of the ACC~counters. Muons are
identified based on their penetration range and transverse scattering
in the KLM~detector. In the momentum region relevant to this analysis,
charged leptons are identified with an efficiency of about 90\% and
the probability to misidentify a pion as an electron (muon) is 0.25\%
(1.4\%)~\cite{Hanagaki:2001fz,Abashian:2002bd}. No SVD hit requirement
is made for lepton tracks. In the lab frame, the (transverse) momentum
of the lepton is required to exceed 0.85 GeV/$c$ (0.6 GeV/$c$). We
also apply an upper lepton momentum cut at 2.4~GeV in the c.m.\ frame
to reject continuum. Again, a vertex fit is performed and
$D^{*+}\ell^-$~candidates are rejected if the vertex probability is
less than $10^{-3}$.

\begin{figure}
  \begin{center}
    \includegraphics[width=0.4\columnwidth]{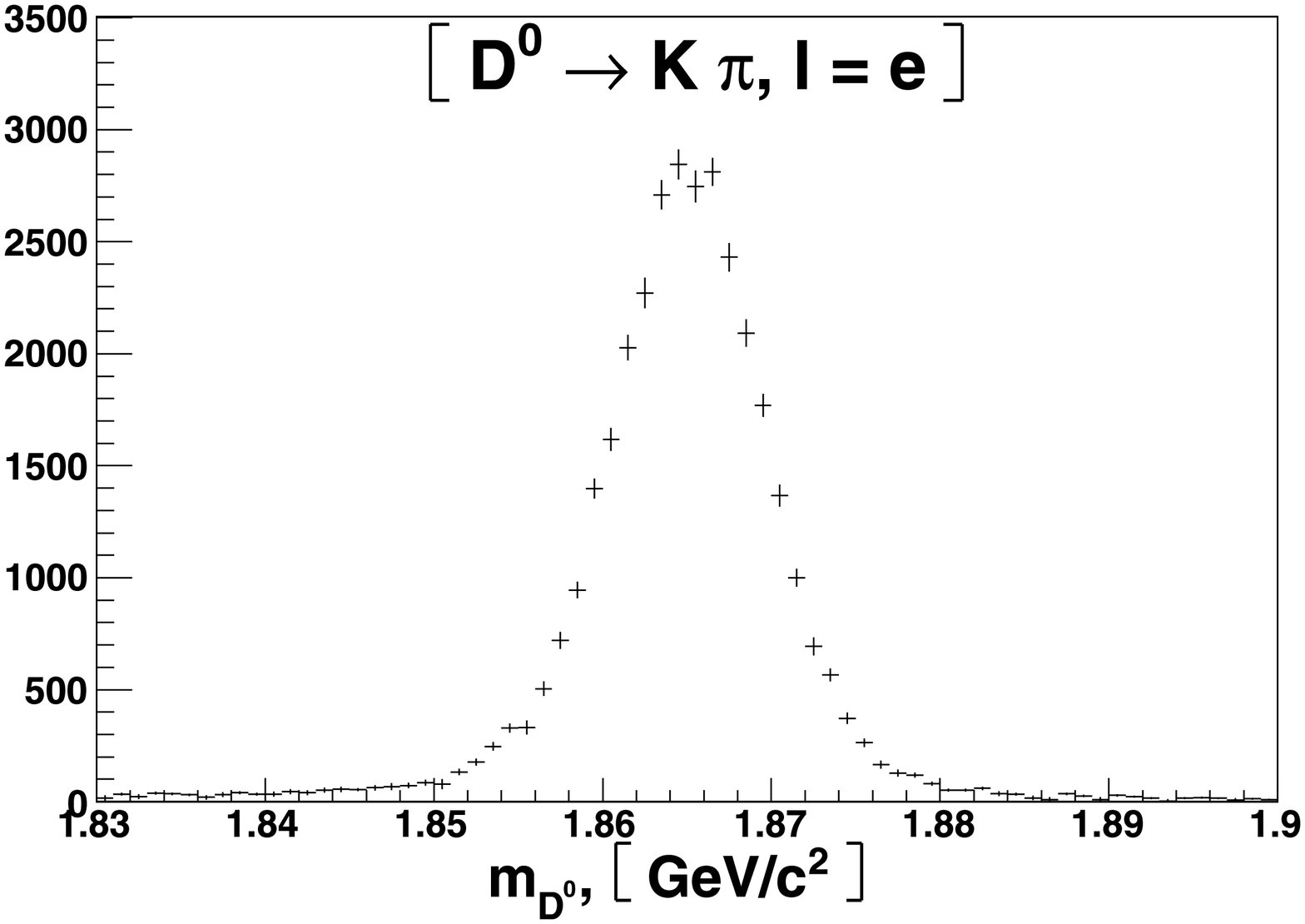}
    \includegraphics[width=0.4\columnwidth]{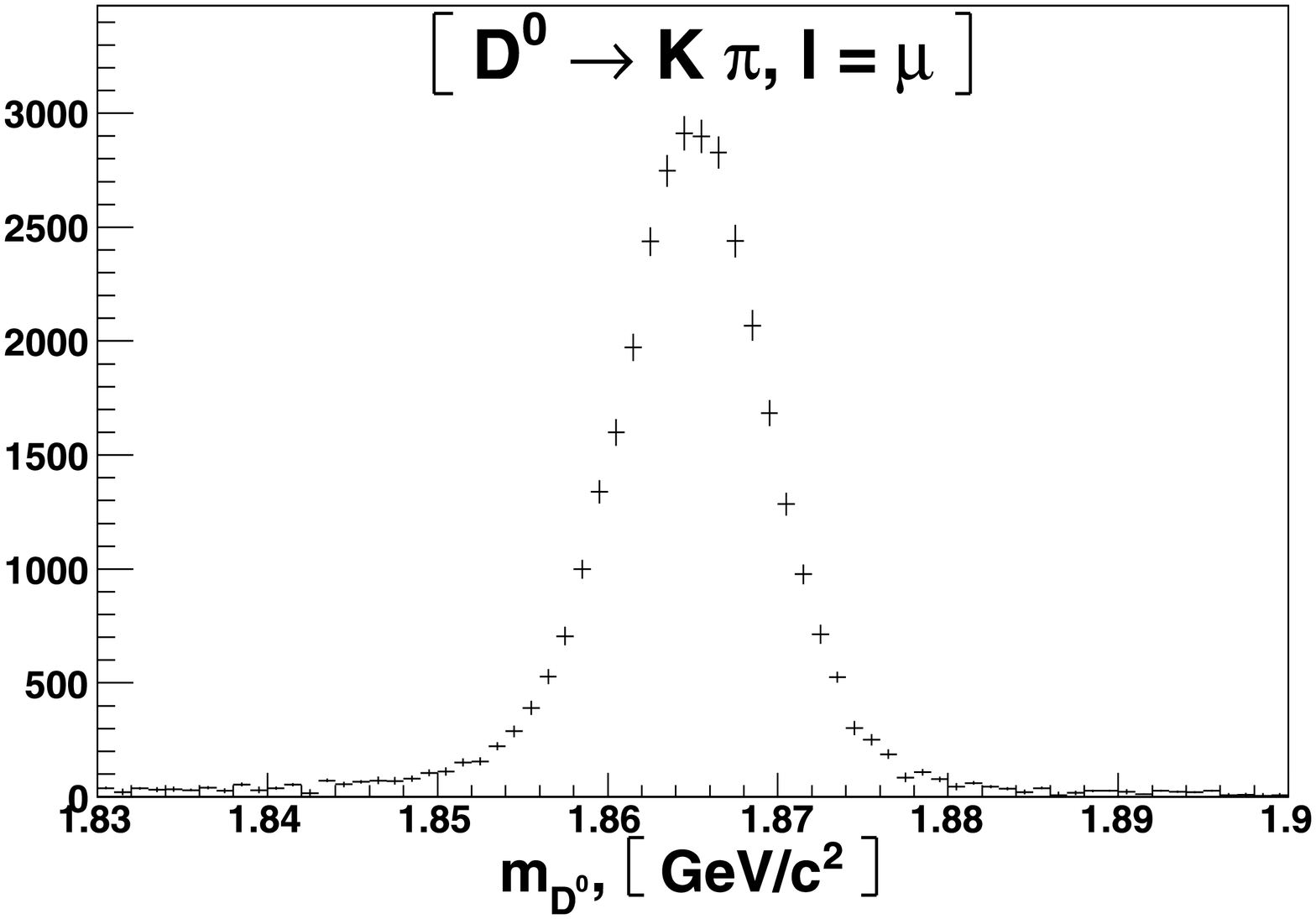}\\
    \includegraphics[width=0.4\columnwidth]{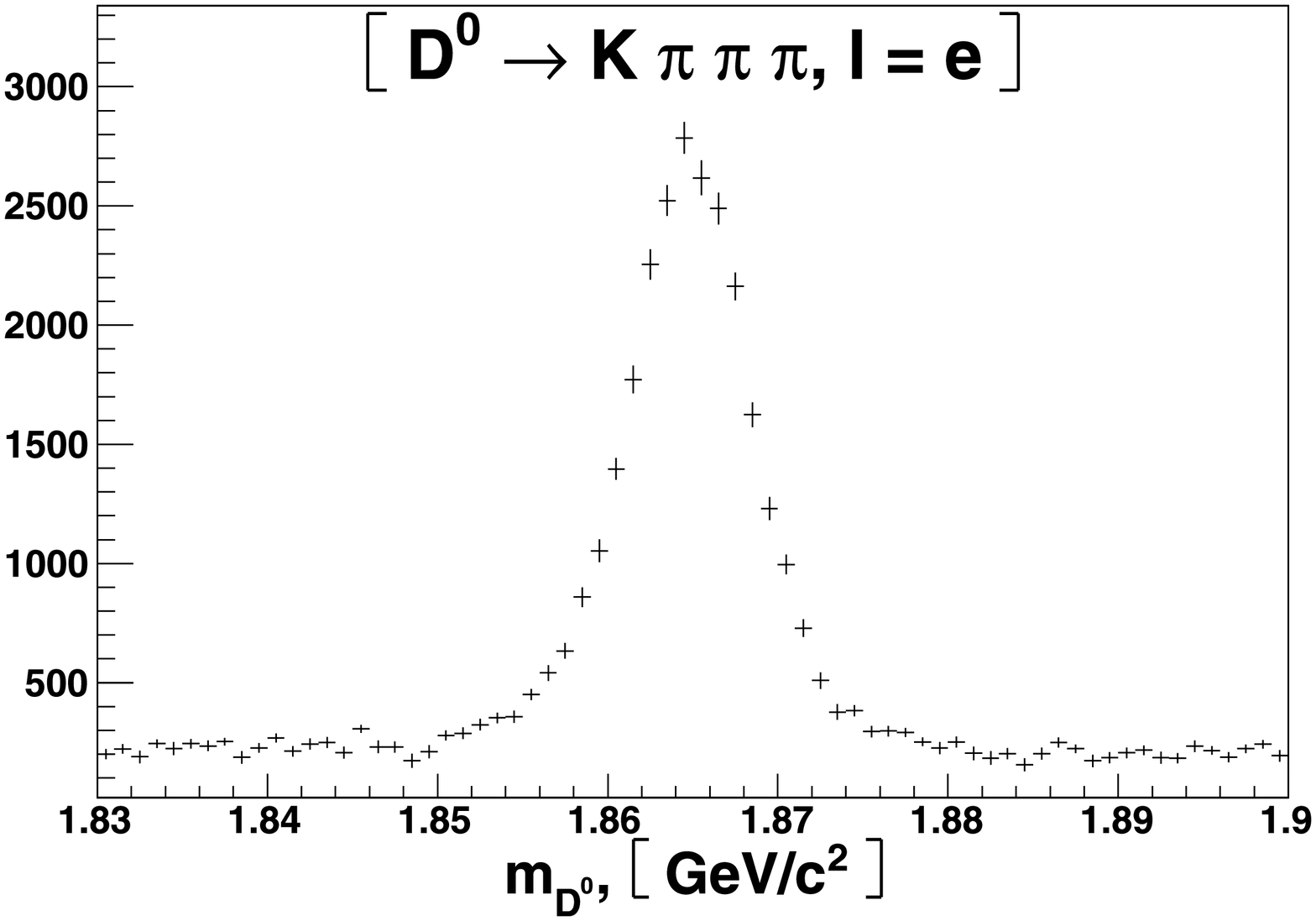}
    \includegraphics[width=0.4\columnwidth]{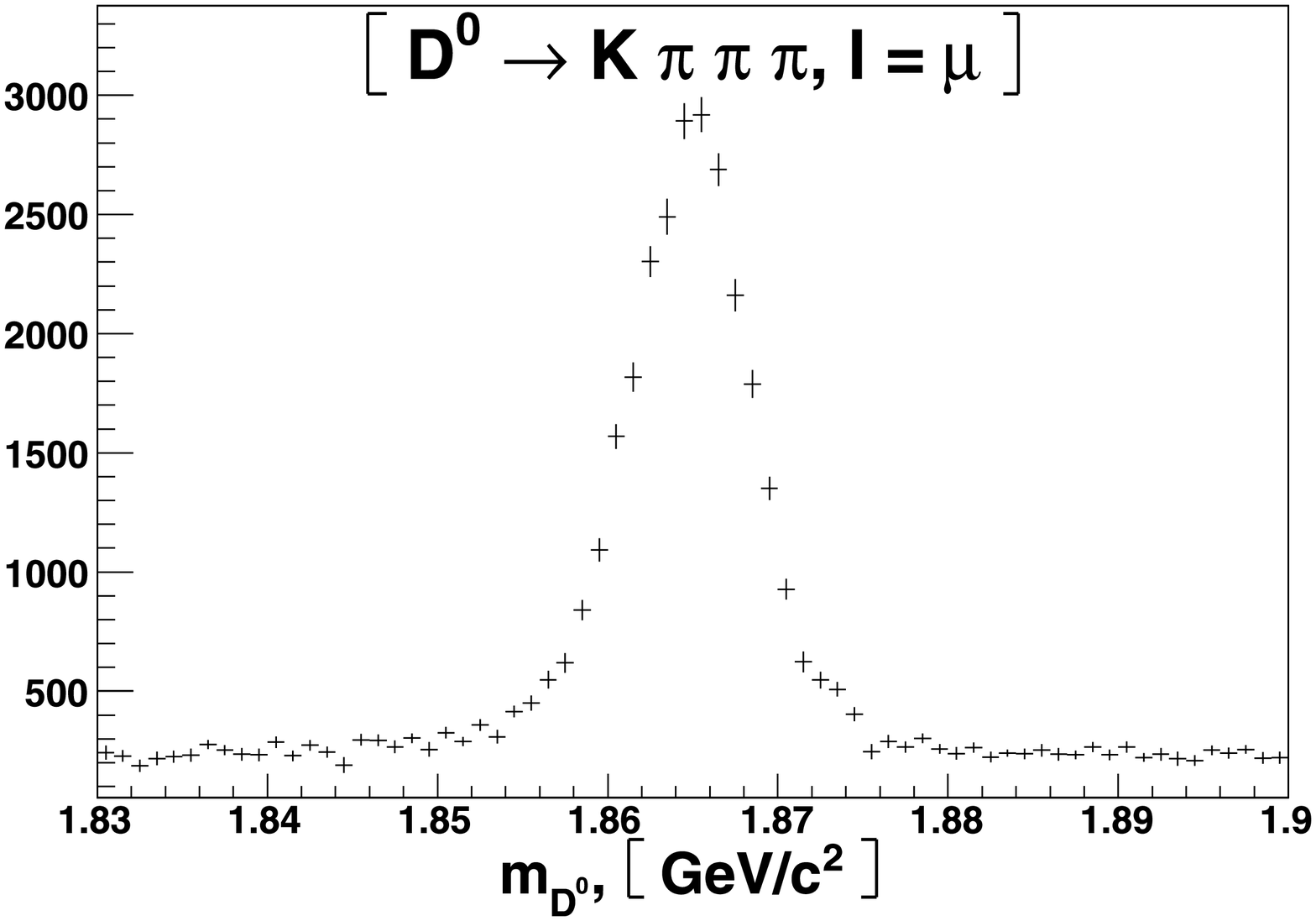}
  \end{center}
  \caption{Invariant $D^0$~candidate mass distributions in the
    different sub-samples. All analysis cuts (except on the plotted
    variable) and $|\cos\theta_{B,D^*\ell}|<1$ are applied.}
    \label{fig:3}
\end{figure}

\begin{figure}
  \begin{center}
    \includegraphics[width=0.4\columnwidth]{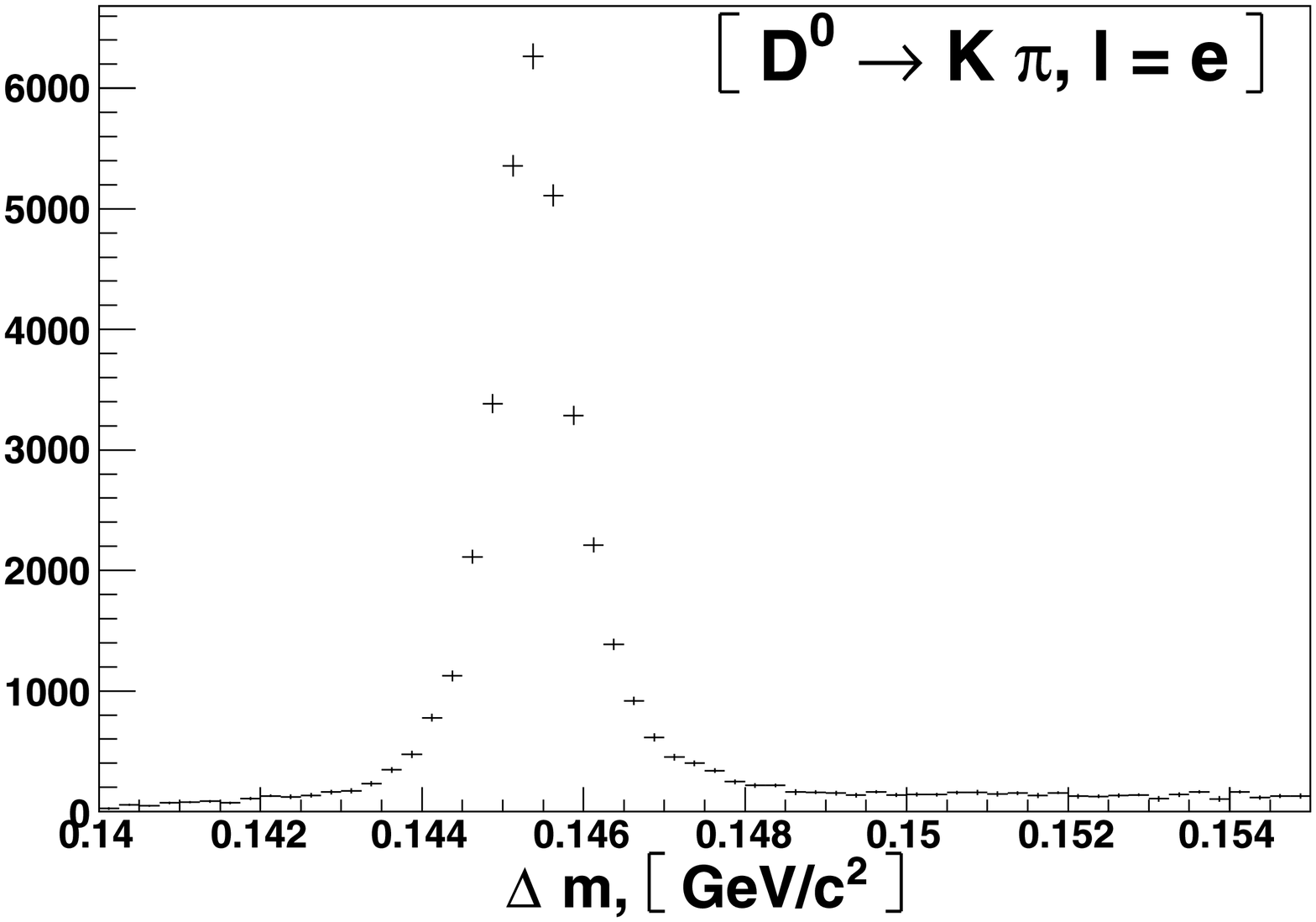}
    \includegraphics[width=0.4\columnwidth]{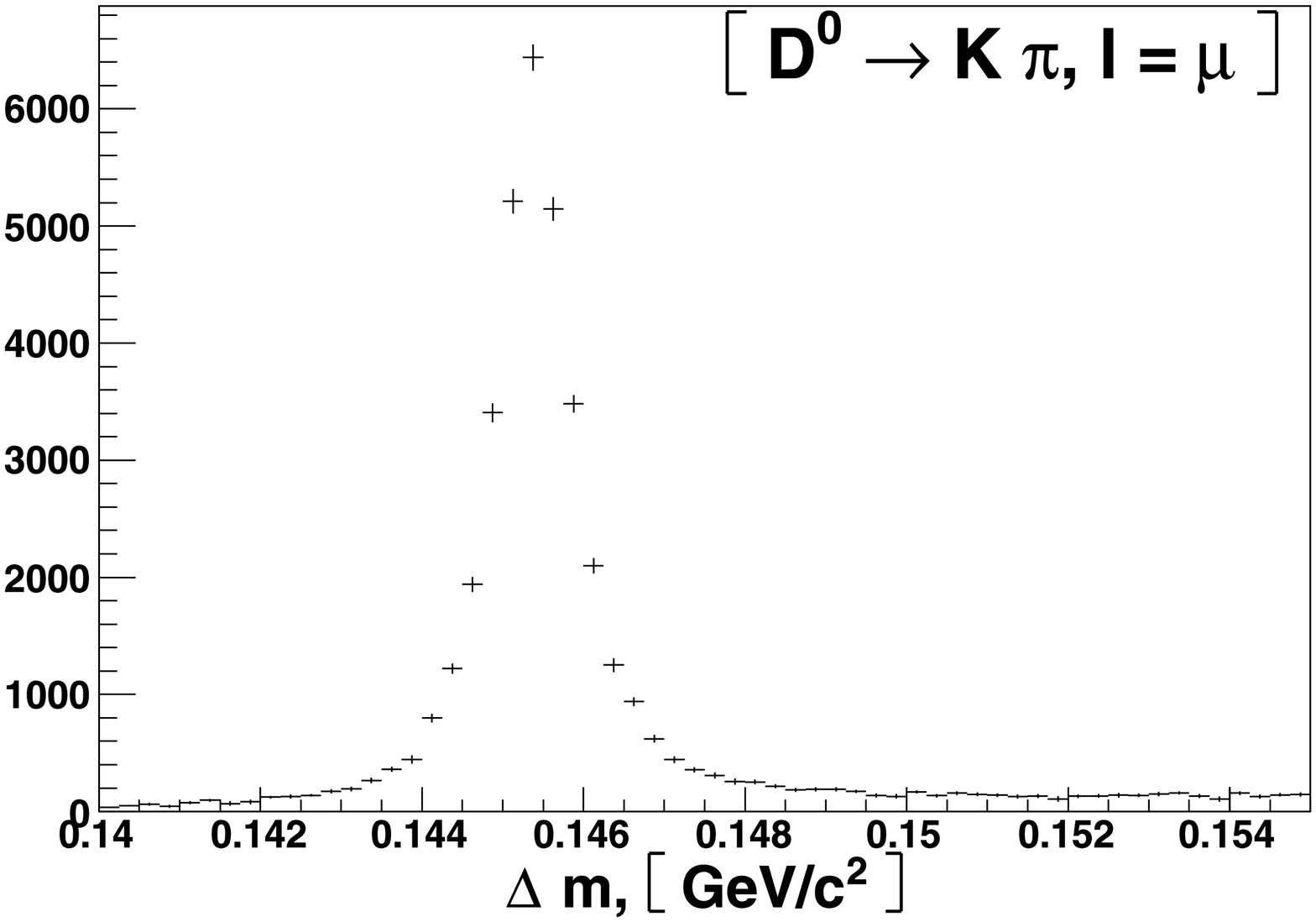}\\
    \includegraphics[width=0.4\columnwidth]{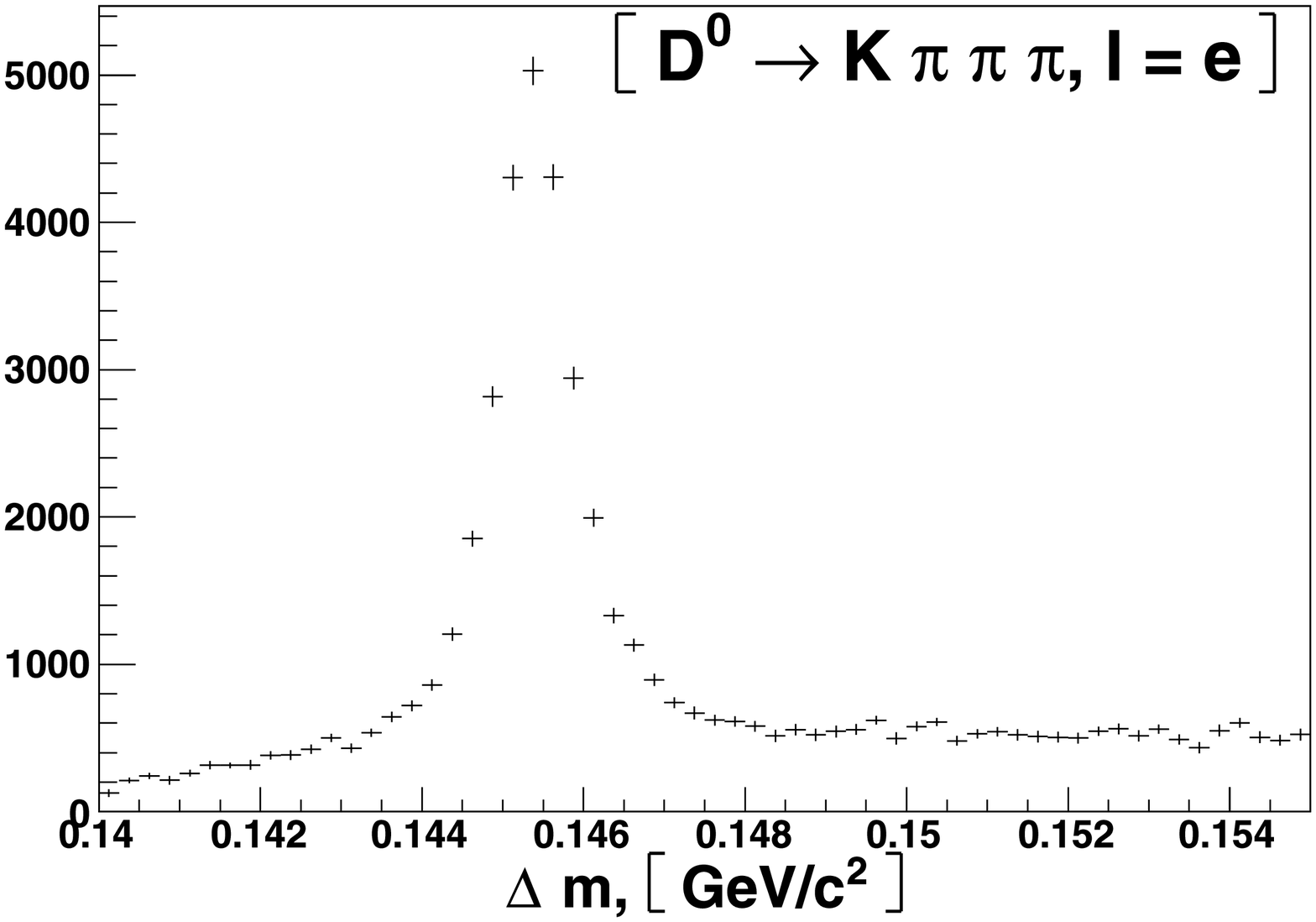}
    \includegraphics[width=0.4\columnwidth]{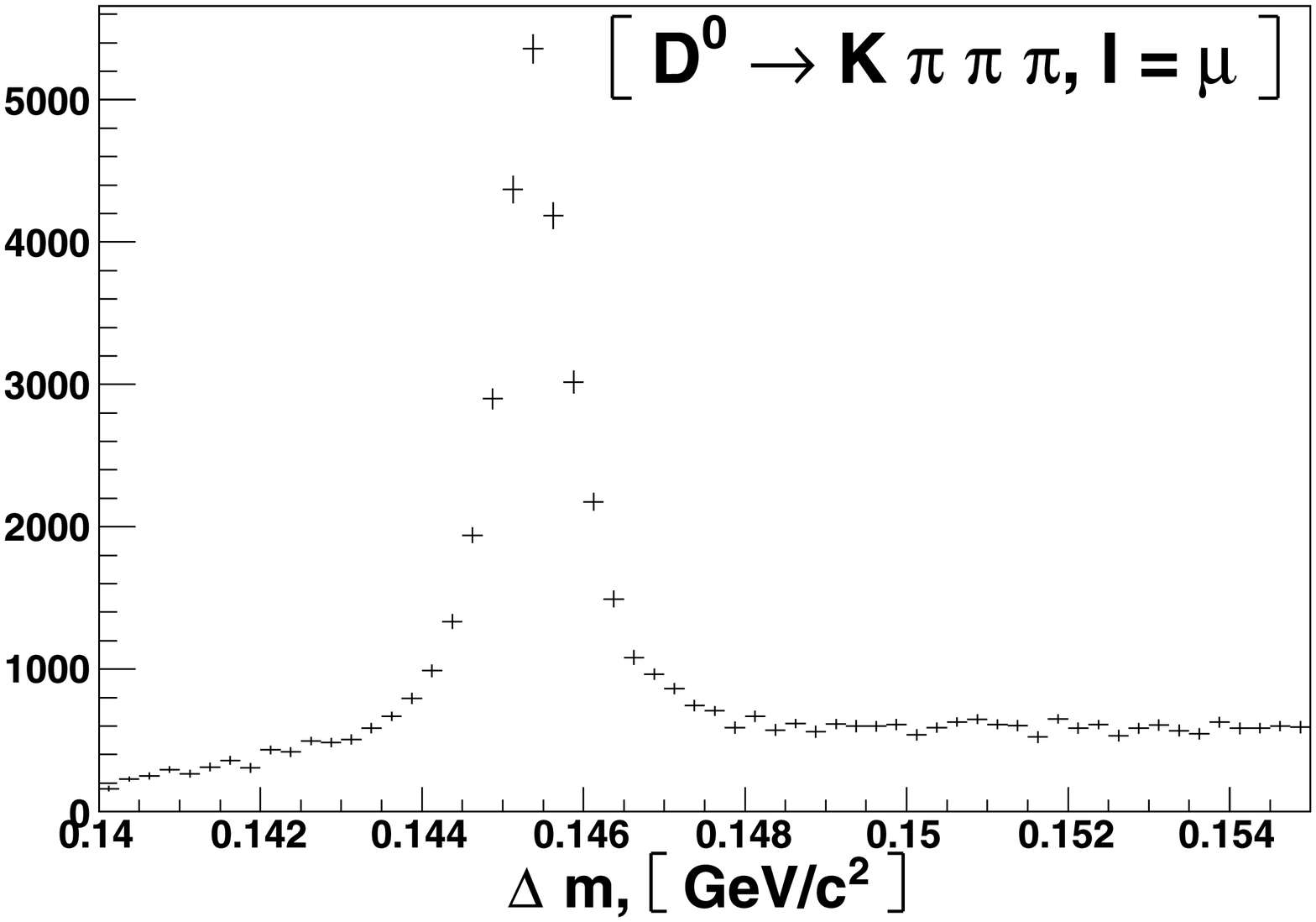}
  \end{center}
  \caption{$\Delta m$~distributions in the different sub-samples. All
  analysis cuts (except on the plotted variable) and
  $|\cos\theta_{B,D^*\ell}|<1$ are applied.} \label{fig:4}
\end{figure}

Figures \ref{fig:3} and \ref{fig:4} show the invariant mass of the
$D^0$~candidates and the $\Delta m$~distributions, respectively.
 
\subsection{Background estimation} \label{sec:3c}

Because we do not reconstruct the other $B$~meson in the event, the
$B$ momentum is \textit{a priori} unknown. However, in the c.m.\
frame, one can show that the $B$~direction lies on a cone around the
$(D^*\ell)$-axis~\cite{ref:2},
\begin{equation}
  \cos\theta_{B,D^*\ell}=\frac{2E^*_BE^*_{D^*\ell}-m^2_B-m^2_{D^*\ell}}{2|\vec
  p^*_B||\vec p^*_{D^*\ell}|}~, \label{eq:3_1}
\end{equation}
where $E^*_B$ is half of the c.m.\ energy and $|\vec p^*_B|$ is
$\sqrt{E^{*2}_B-m^2_B}$. The quantities $E^*_{D^*\ell}$, $\vec
p^*_{D^*\ell}$ and $m_{D^*\ell}$ are calculated from the reconstructed
$D^*\ell$~system.

This cosine is also a powerful discriminator between signal and
background: Signal events should strictly lie in the interval $(-1,1)$,
although -- due to finite detector resolution -- about 8\% of the signal is
reconstructed outside this interval. The background on the other hand
does not have this restriction. We therefore perform a fit to the
$\cos\theta_{B,D^*\ell}$~distribution to determine the background
normalizations from the data.

The background contained in the selected events can be attributed to
the following six sources:
\begin{itemize}
\item continuum: any candidate reconstructed in a non-$\Upsilon(4S)$
  event

\item fake lepton: the charged lepton candidate is fake; the
    $D^*$~candidate might be fake or not

\item fake $D^*$: the $D^*$~candidate is misreconstructed; the lepton
    candidate is an actual electron or muon

\item $D^{**}$: background from $B\to\bar D^{**}\ell^+\nu$~decays with
    $\bar D^{**}\to D^{*-}\pi$ or $B\to D^{*-}\pi\ell^+\nu$
    non-resonant

\item correlated background: background from other processes in
    which the $D^*$ and the lepton stem from the same $B$~meson, {\it e.g},
    $B^0\to D^{*-}\tau^+\nu$, $\tau^+\to\mu^+\nu\nu$

\item uncorrelated background: the $D^*$ and the lepton stem from
  different $B$~mesons
\end{itemize}

\begin{figure}
  \begin{center}
    \includegraphics[width=0.4\columnwidth]{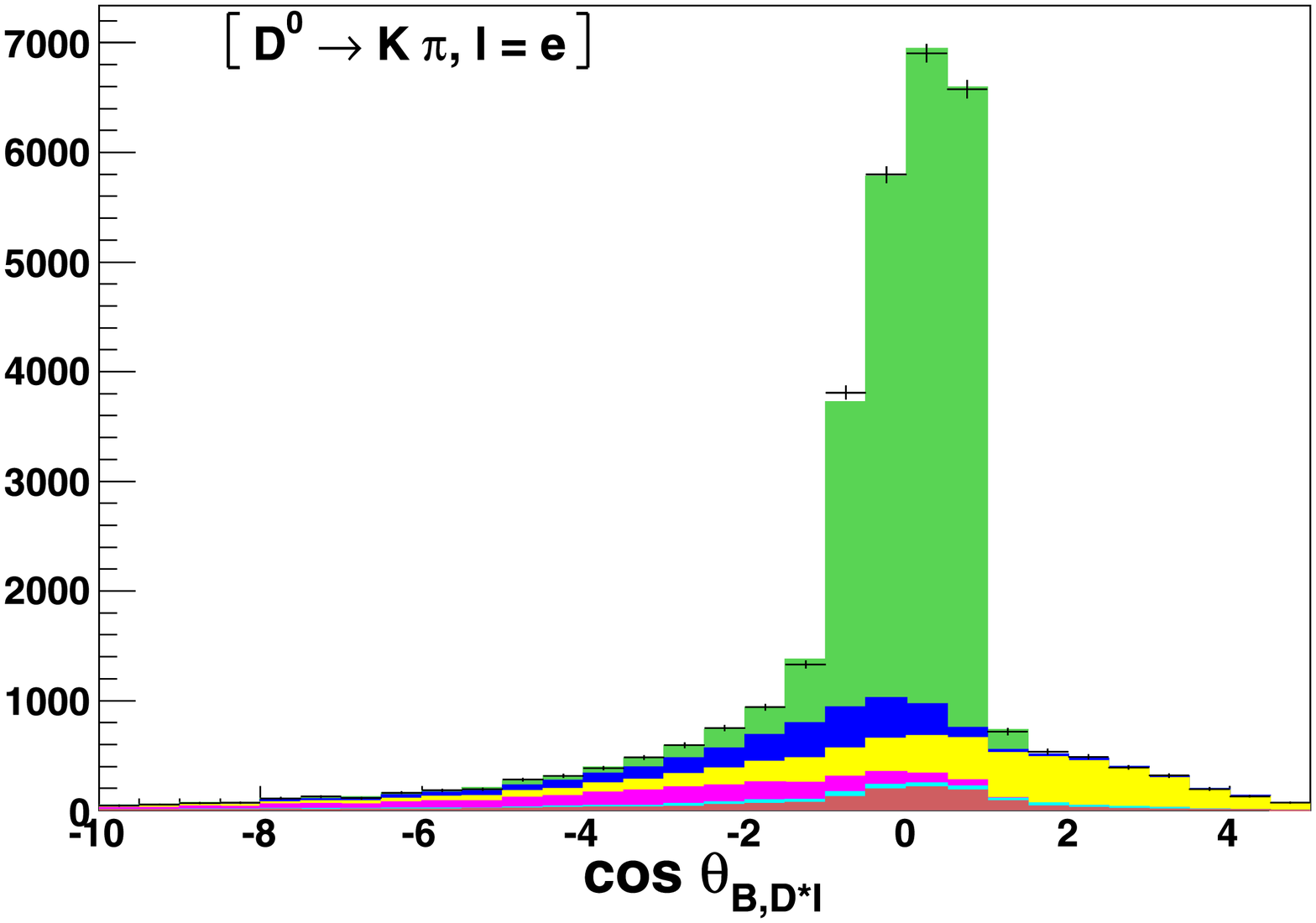}
    \includegraphics[width=0.4\columnwidth]{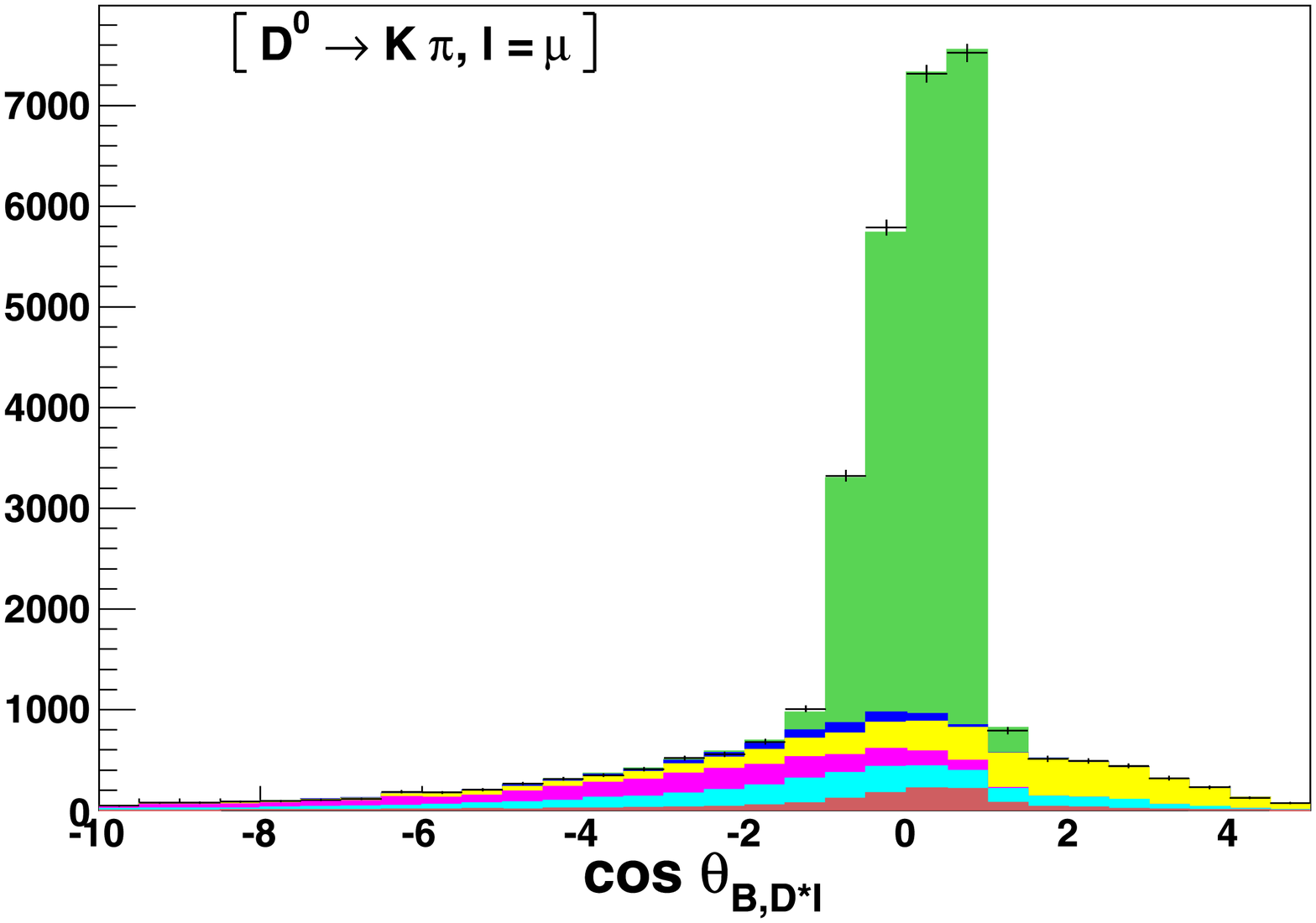}\\
    \includegraphics[width=0.4\columnwidth]{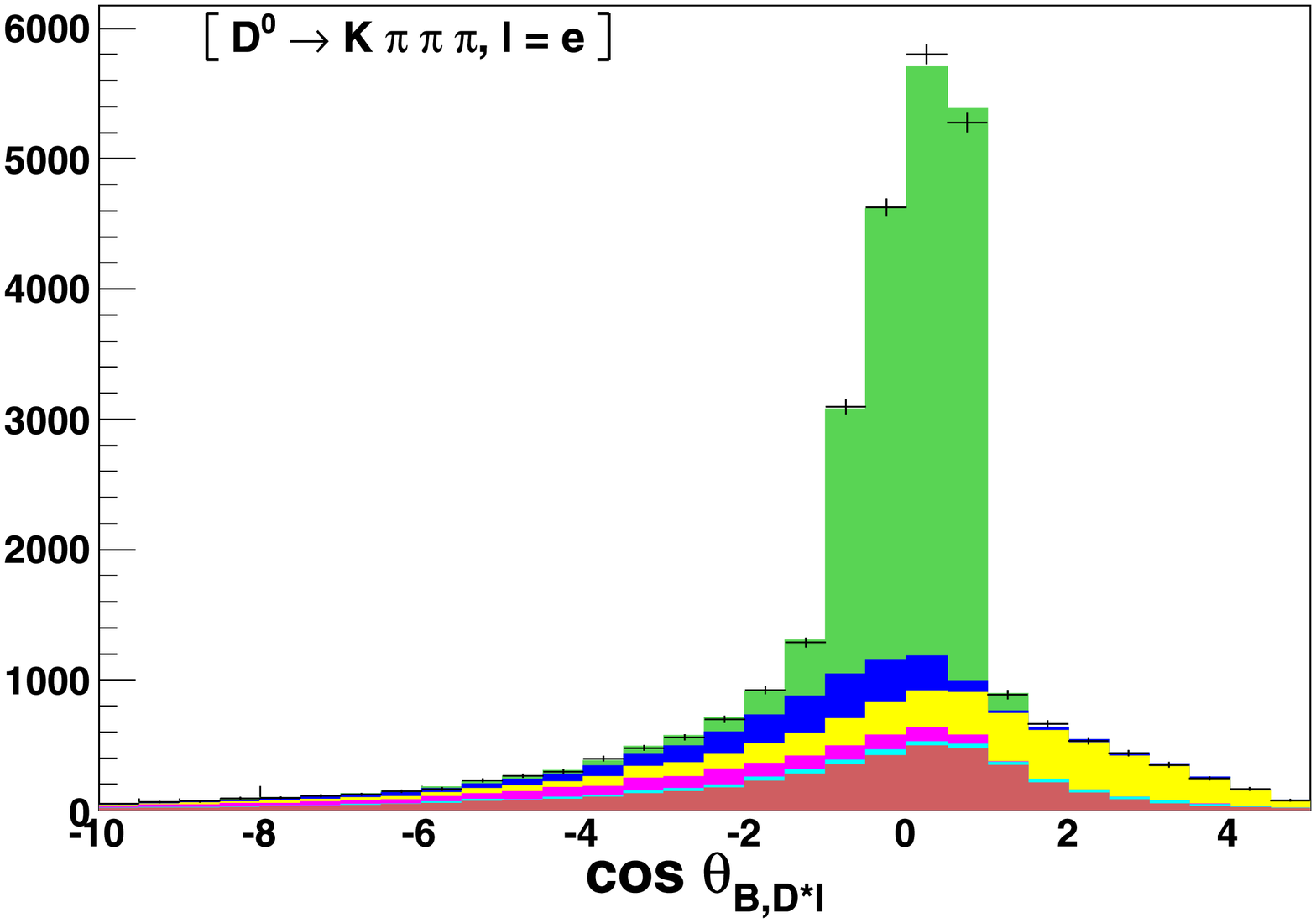}
    \includegraphics[width=0.4\columnwidth]{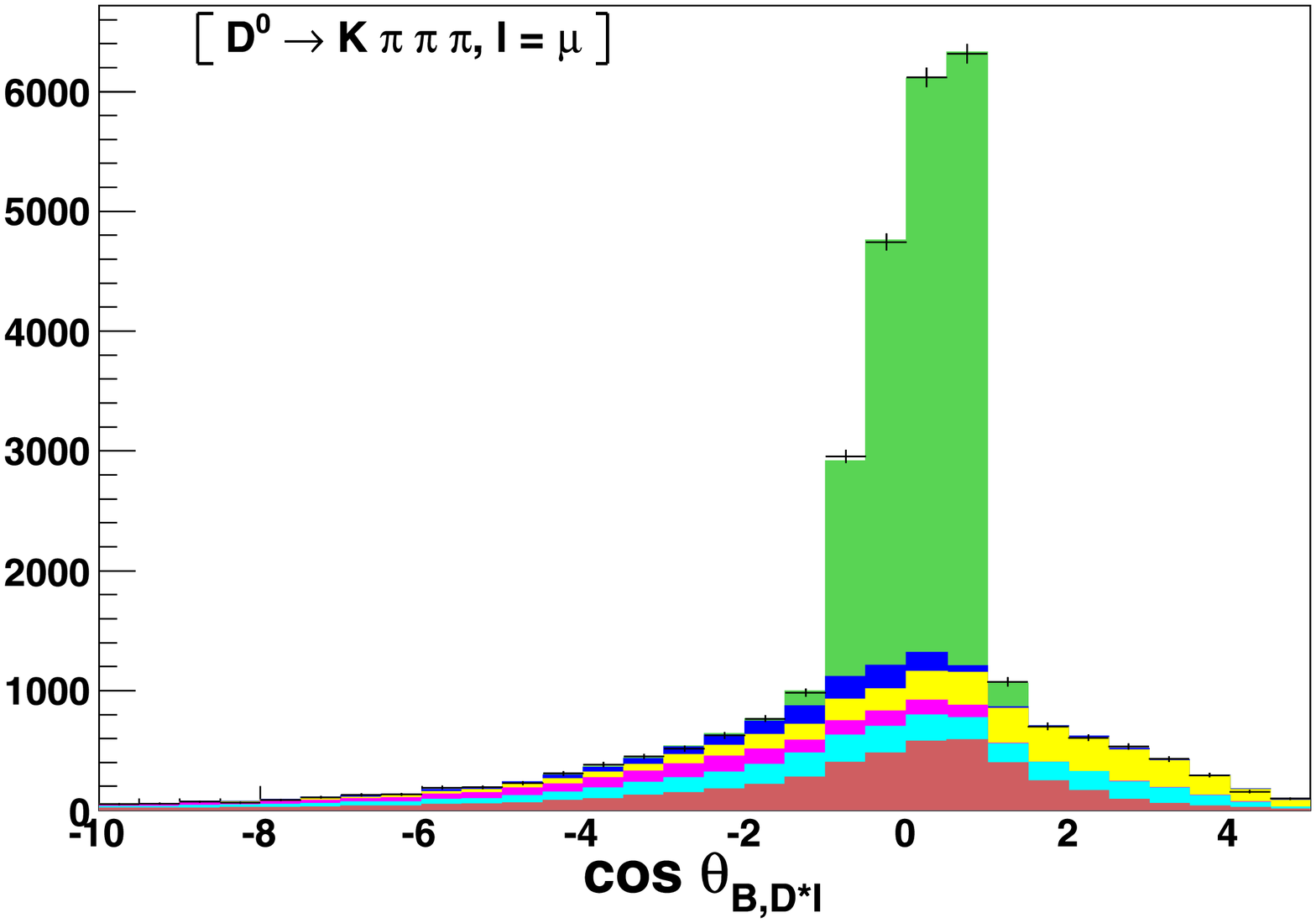}\\
    \includegraphics[width=0.4\columnwidth]{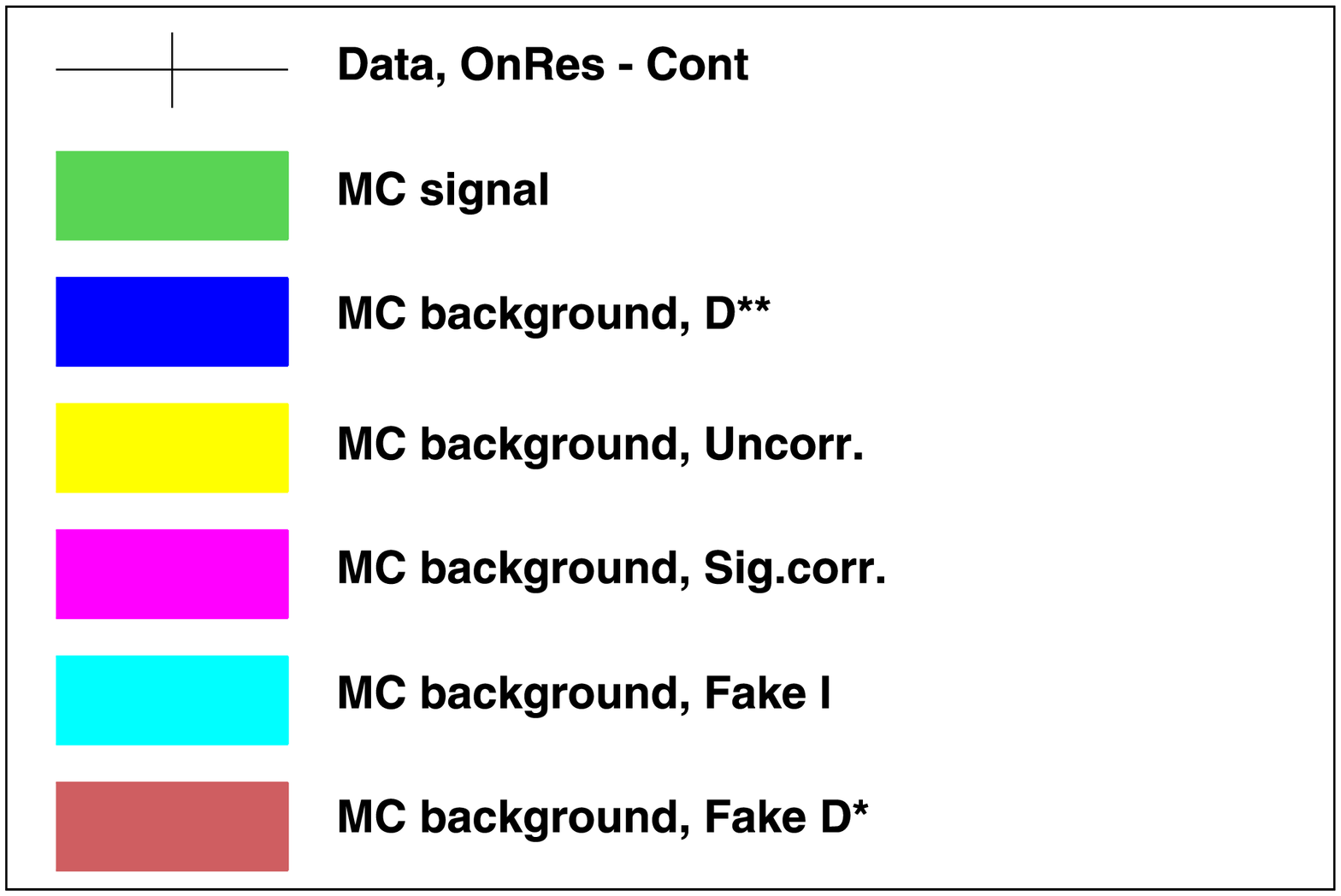}
  \end{center}
  \caption{Result of the fits to the
    $\cos\theta_{B,D^*\ell}$~distributions in the different
    sub-samples.} \label{fig:5}.
\end{figure}

These background components are modeled by $(D^*\ell)$-candidates,
appropriately selected from MC data based on generator information,
except continuum which is modeled by off-resonance events. The shape
of the fake muon background is corrected by the ratio of the pion fake
rate in the experimental data over the same quantity in the MC
simulation, as measured using kinematically identified pions in
$K^0_S\to\pi^+\pi^-$~decays. The $\cos\theta_{B,D^*\ell}$~distribution
in the data is fitted using the {\tt TFractionFitter}
algorithm~\cite{Barlow:1993dm} in {\tt ROOT}~\cite{Brun:1997pa}. The
fit is done separately in the four sub-samples defined by the $D^0$
decay channel and the lepton type. The results are shown in
Fig.~\ref{fig:5} and Table~\ref{tab:1}.
\begin{table}
  \begin{center}
    \begin{tabular}{l|@{\extracolsep{.3cm}}cccc}
      \hline\hline
      sample & $K\pi,e$ & $K\pi,\mu$ & $K3\pi,e$ & $K3\pi,\mu$\\
      \hline
      signal & $(80.95\pm 1.06)$\% & $(80.92\pm 0.98)$\% & $(73.17\pm
      1.71)$\% & $(72.22\pm 1.46)$\%\\
      $D^{**}$ & $(4.73\pm 0.87)$\% & $(1.24\pm 0.85)$\% &
      $(5.21\pm1.18)$\% & $(2.85\pm 1.10)$\%\\
      uncorrelated & $(5.36\pm 0.27)$\% & $(4.38\pm 0.29)$\%&$(5.42\pm
      0.58)$\% & $(4.17\pm 0.54)$\%\\
      correlated & $(1.69\pm 0.26)$\% & $(2.42\pm 0.28)$\% & $(2.04\pm
      0.69)$\% & $(2.25\pm 0.59)$\%\\
      fake $\ell$ & 0.68\% \scriptsize{(fixed)} & 3.62\%
      \scriptsize{(fixed)} & 0.72\% \scriptsize{(fixed)} &
      4.04\%\scriptsize{(fixed)}\\
      fake $D^{*}$ & 2.96\% \scriptsize{(fixed)} & 2.91\%
      \scriptsize{(fixed)} & $(8.78\pm 2.63)$\% & $(9.63\pm 2.15)$\%\\
      continuum & 3.62\% \scriptsize{(fixed)} & 4.51\%
      \scriptsize{(fixed)} & 4.81\% \scriptsize{(fixed)} & 4.87\%
      \scriptsize{(fixed)}\\
      \hline\hline
    \end{tabular}
  \end{center}
  \caption{The signal and background fractions for selected events
    within the signal window $|\cos\theta_{B,D^*\ell}|<1$.} \label{tab:1}
\end{table}

In all fits, the continuum normalization is fixed to the on- to
off-resonance luminosity ratio, corrected for the $1/s$~dependence of
the $e^+e^-\to q\bar q$ cross-section. The normalizations of some
other components have also been fixed to the MC expectations if their
contributions cannot be determined reliably from the
$\cos\theta_{B,D^*\ell}$~spectrum, as indicated in the table. In
general, the normalizations obtained by the fit agree well with the MC
expectations except for the $D^{**}$~component which is too abundant
in the MC.

\subsection{Kinematic variables}

\begin{figure}
  \begin{center}
    \includegraphics[width=0.8\columnwidth]{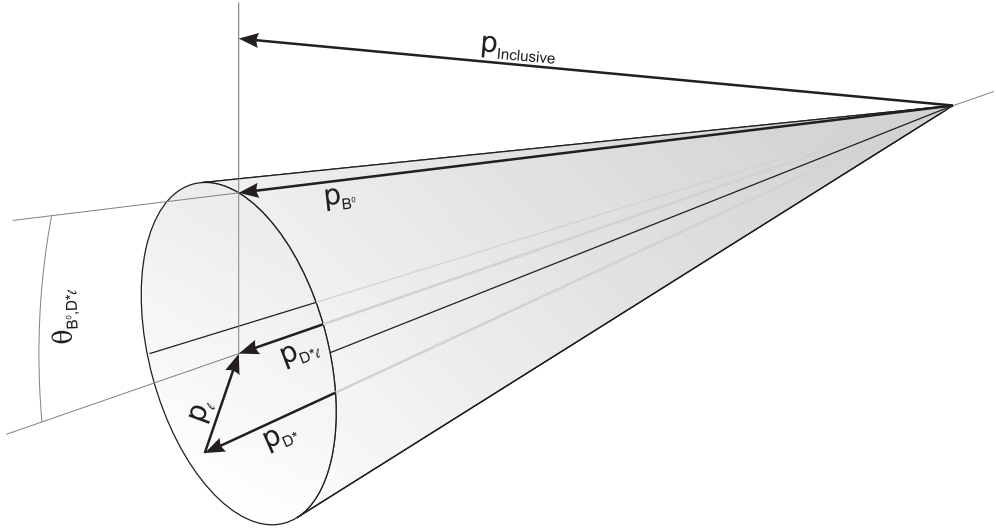}
  \end{center}
  \caption{Reconstruction of the $B^0$~direction. Refer to the text
    for details.} \label{fig:6}
\end{figure}

To calculate the four kinematic variables -- $w$, $\cos\theta_\ell$,
$\cos\theta_V$ and $\chi$ -- that characterize the $B^0\to
D^{*-}l^+\nu$~decay defined in Sect.~\ref{sec:2a}, we need to
determine the $B^0$~rest frame. The $B$~direction is already known to
be on a cone around the $(D^*\ell)$-axis with opening angle
$2\theta_{B,D^*\ell}$ in the c.m.\ frame, Eq.~(\ref{eq:3_1}). For the best
guess of the $B$~direction, we first estimate the c.m.\ frame
$B$~vector by summing the momenta of the remaining particles in the
event ($\vec p^*_\mathrm{inclusive}$~\cite{ref:2}) and choose the
direction on the cone that minimizes the difference to $\vec
p^*_\mathrm{inclusive}$, as shown in Fig.~\ref{fig:6}.

To obtain $\vec p^*_\mathrm{inclusive}$, we exclude tracks
passing very far away from the interaction point or compatible with a
multiply reconstructed track generated by a low-momentum particle
spiraling in the central drift chamber. Unmatched clusters in the
barrel region must have an energy greater than 50~MeV. For clusters in
the forward (backward) region, the threshold is at 100~MeV
(150~MeV). Then, we compute $\vec p_\mathrm{inclusive}$ (in the lab.\
frame) by summing the 3-momenta of the selected particles,
\begin{equation}
  \vec p_\mathrm{inclusive}=\vec p_\mathrm{HER}+\vec
  p_\mathrm{LER}-\sum_i \vec p_i~,
\end{equation}
where the indices HER and LER correspond to the colliding beams, and
transform this vector into the c.m.\ frame. Note that we do not make
any mass assumption for the charged particles. The energy component of
$p_\mathrm{inclusive}$ is defined by requiring
$E^*_\mathrm{inclusive}$ to be $E^*_\mathrm{beam}=\sqrt{s}/2$.

With the $B^0$~rest frame reconstructed in this way, the resolutions
in the kinematic variables are found to be about 0.025, 0.052, 0.047
and 6.47$^\circ$ for $w$, $\cos\theta_\ell$, $\cos\theta_V$ and
$\chi$, respectively.

\subsection{Fit procedure}

In the rest of this analysis, we consider only events passing the
signal window requirement $|\cos\theta_{B,D^*\ell}|<1$. We perform a
binned $\chi^{2}$~fit of the $w$, $\cos\theta_\ell$, $\cos\theta_V$
and $\chi$~distributions over (almost) the entire phase space to
measure the following quantities: the form factor normalization
$\mathcal{F}(1)|V_{cb}|$ Eq.~(\ref{eq:2_2}), and the three parameters
$\rho^2$, $R_1(1)$ and $R_2(1)$ which parameterize the form factor in
the HQET framework Eqs.~(\ref{eq:2_3})--(\ref{eq:2_5}). Instead of fitting
in four dimensions, we fit simultaneously the one-dimensional
projections of $w$, $\cos\theta_\ell$, $\cos\theta_V$ and $\chi$ to
have enough entries in each bin of the fit. This introduces bin-to-bin
correlations which have to be accounted for.

The distributions in $w$, $\cos\theta_\ell$, $\cos\theta_V$ and $\chi$
are divided into ten bins of equal width. The kinematically allowed
values of $w$ are between $1$ and $\approx 1.504$ but we restrict the
$w$~range to values between $1$ and $1.5$. In each sub-sample, there
are thus 40~bins to be used in the fit. In the following, we label
these bins with a common index $i$, $i=1,\dots,40$, \textit{i.e.},
depending on the value of $i$, the $i^\mathrm{th}$~bin might belong to
the $w$,  $\cos\theta_\ell$, $\cos\theta_V$ or $\chi$~distribution.

The predicted number of events $N^\mathrm{th}_i$ in the bin~$i$ is
given by
\begin{equation}
  N^\mathrm{th}_i=N_{B^0}\mathcal{B}(D^{*+}\to
  D^0\pi^+)\mathcal{B}(D^0)\tau_{B^0}\Gamma_i~,
\end{equation}
where $N_{B^0}$ is the number of $B^0$~mesons in the data sample and
$\mathcal{B}(D^{*+}\to D^0\pi^+)$ is taken from
Ref.~\cite{Yao:2006px}. In the $(K\pi,e)$ and $(K\pi,\mu)$
sub-samples $\mathcal{B}(D^0)$ is $\mathcal{B}(D^0\to
K^-\pi^+)$~\cite{Yao:2006px}; in $(K3\pi,e)$ and $(K3\pi,\mu)$
$\mathcal{B}(D^0)$ is $R_{K3\pi/K\pi}\mathcal{B}(D^0\to K^-\pi^+)$,
with $R_{K3\pi/K\pi}$ a fifth free parameter of the fit. Finally,
$\tau_{B^0}$ is the $B^0$~lifetime~\cite{Yao:2006px}, and $\Gamma_i$
is the width obtained by integrating Eq.~(\ref{eq:2_1}) in the kinematic
variable corresponding to $i$ from the lower to the upper bin boundary
(the other kinematic variables are integrated over their full range). This
integration is numerical in the case of $w$ and analytic for the other
variables. The expected number of events $N^\mathrm{exp}_i$ is related
to $N^\mathrm{th}_i$ as follows
\begin{equation}
  N^\mathrm{exp}_i=\sum_{j=1}^{40}\left(R_{ij}\epsilon_jN^\mathrm{th}_j\right)+N^\mathrm{bkg}_i~.
\end{equation}
Here, $\epsilon_i$ is the probability that an event generated in the
bin~$i$ is reconstructed and passes all analysis cuts, and $R_{ij}$ is the
detector response matrix, \textit{i.e.}, it gives the probability that
an event generated in the bin~$j$ is observed in the bin~$i$. Both
quantities are calculated using MC~simulation. $N^\mathrm{bkg}_i$ is
the number of expected background events, estimated as described in
Sect.~\ref{sec:3c}.

Next, we calculate the variance~$\sigma^2_i$ of $N^\mathrm{exp}_i$. We
consider the following contributions: the uncertainty in
$N^\mathrm{th}_i$ (poissonian); fluctuations related to $N$
repetitions of the yes/no experiment with known success probability
$\epsilon_i$ (binomial); a similar contribution to the variance
related to $R_{ij}$ (multinomial); and the uncertainty in the
background contribution $N^\mathrm{bkg}_i$. This yields the
following expression for $\sigma^2_i$,
\begin{equation}
  \begin{split}
    \sigma^2_i=\sum_{j=1}^{40}\left[R^2_{ij}\epsilon^2_jN^\mathrm{th}_j+R^2_{ij}\frac{\epsilon_j(1-\epsilon_j)}{N_\mathrm{data}}(N^\mathrm{th}_j)^2+\frac{R_{ij}(1-R_{ij})}{N'_\mathrm{data}}\epsilon^2_j(N^\mathrm{th}_j)^2+\right.\\
      \left.R^2_{ij}\frac{\epsilon_j(1-\epsilon_j)}{N_\mathrm{MC}}(N^\mathrm{th}_j)^2+\frac{R_{ij}(1-R_{ij})}{N'_\mathrm{MC}}\epsilon^2_j(N^\mathrm{th}_j)^2\right]+\sigma^2(N^\mathrm{bkg}_i)~.
  \end{split}
\end{equation}
The first term is the poissonian uncertainty in $N^\mathrm{th}_i$. The
second and third terms are the binomial and multinomial uncertainties
related to the finite real data size, respectively, where
$N_\mathrm{data}$ ($N'_\mathrm{data}$) is the total number of decays
(the number of reconstructed decays) into the final state under
consideration ($K\pi$ or $K3\pi$, $e$ or $\mu$) in the real data. The
quantities $\epsilon_i$ and $R_{ij}$ are calculated from a finite
signal MC sample ($N_\mathrm{MC}$ and $N'_\mathrm{MC}$); the corresponding
uncertainties are estimated by the fourth and fifth terms. Finally,
the last term is the background contribution
$\sigma^2(N^\mathrm{bkg}_i)$, calculated as the sum of the different
background component variances. For each background component
defined in Sect.~\ref{sec:3c} we estimate its contribution by linear error
propagation of the scale factor and the error determined by the
procedure described above. For continuum, we estimate the error in the
on-resonance to off-resonance luminosity ratio to be 1.5\%.

In each sub-sample. we calculate the off-diagonal elements of the
covariance matrix~cov$_{ij}$ as $Np_{ij}-Np_ip_j$, where $p_{ij}$ is
the relative abundance of the bin~$(i,j)$ in the 2-dimensional histogram
obtained by plotting the kinematic variables against each other,
$p_i$ is the relative number of entries in the 1-dimensional
distribution, and $N$ is the size of the sample. Covariances are
calculated for the signal and the different background components, and
added with appropriate normalizations.

The covariance matrix is inverted numerically within {\tt ROOT} and,
labelling the four sub-samples ($K\pi$ or $K3\pi$, $e$ or $\mu$) with
the index~$k$, the sub-sample $\chi^2$~functions are calculated,
\begin{equation}
  \chi^2_k=\sum_{i,j}(N^\mathrm{obs}_i-N^\mathrm{exp}_i)C^{-1}_{ij}(N^\mathrm{obs}_j-N^\mathrm{exp}_j)~,
\end{equation}
where $N^\mathrm{obs}_i$ is the number of events observed in bin~$i$
in the data. We sum these four functions and minimize the global
$\chi^2$ with {\tt MINUIT}~\cite{James:1975dr}.

We have tested this fit procedure using generic MC data
samples. All results are consistent with expectations and show no
indication of bias.

\section{Results and systematic uncertainties}

\subsection{Results}

\begin{figure}
  \begin{center}
    \includegraphics[width=0.8\columnwidth]{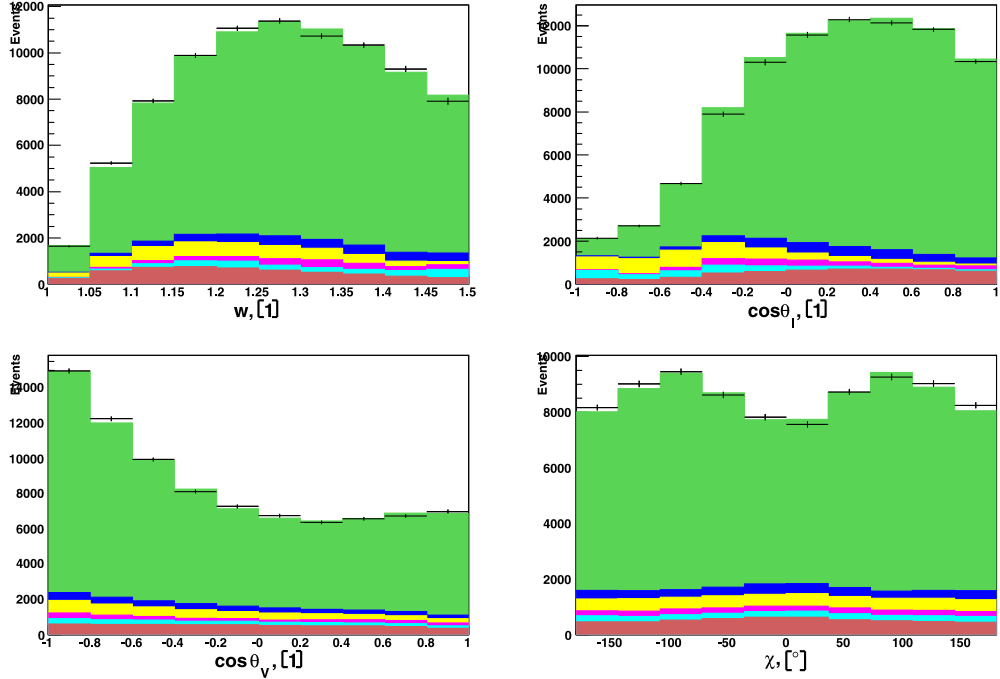}
  \end{center}
  \caption{Preliminary result of the fit of the four kinematic
    variables in the total sample. (The different sub-samples are
    added in this plot.) The points with error bars are continuum
    subtracted on-resonance data. The histograms are the signal and
    the different background components. The color scheme is explained
    in Fig.~\ref{fig:5}.} \label{fig:7}
\end{figure}

After applying all selection requirements and subtracting backgrounds,
$69,345\pm 377$ signal events are found in the data. The preliminary
result of the fit to these events is shown in Fig.~\ref{fig:7} and
Table~\ref{tab:4}. The statistical correlation coefficients of the
five fit parameters are given in Table~\ref{tab:5}.
\begin{table}
  \begin{center}
    \begin{tabular}{l|@{\extracolsep{.3cm}}ccc}
      \hline\hline
      sample & $K\pi,e$ & $K\pi,\mu$ & $K3\pi,e$\\
      \hline
      $\rho^2$ & $1.329\pm 0.072\pm 0.017$ & $1.221\pm 0.075\pm 0.046$
      & $1.238\pm 0.233\pm 0.053$\\
      $R_1(1)$ & $1.455\pm 0.077\pm 0.046$ & $1.608\pm 0.087\pm
      0.099$ & $1.085\pm 0.125\pm 0.044$\\
      $R_2(1)$ & $0.782\pm 0.055\pm 0.014$ & $0.853\pm 0.055\pm
      0.027$ & $0.980\pm 0.087\pm 0.027$\\
      $R_{K3\pi/K\pi}$ & 2.153 \scriptsize{(fixed)} & 2.153
      \scriptsize{(fixed)} & 2.153 \scriptsize{(fixed)}\\
      $\mathcal{B}(B^0\to D^{*-}\ell^+\nu_\ell)$ (\%) & $4.43\pm
      0.03\pm 0.25$ & $4.41\pm 0.03\pm 0.26$ & $4.42\pm 0.04\pm
      0.25$\\
      $\mathcal{F}(1)|V_{cb}|$ (10$^{-3}$) & $34.3\pm 0.4\pm 1.0$ &
      $33.5\pm 0.4\pm 1.0$ & $35.6\pm 0.8\pm 1.3$\\
      $\chi^2/$n.d.f. & 29.2/36 & 37.4/36 & 19.2/36\\
      $P(\chi^2)$ & 78.2\% & 40.4\% & 99.0\%\\
      \hline\hline
    \end{tabular}

    \vspace{.3cm}

    \begin{tabular}{l|@{\extracolsep{.3cm}}ccc}
      \hline\hline
      sample & $K3\pi,\mu$ & average & full sample\\
      \hline
      $\rho^2$ & $1.436\pm 0.121\pm 0.062$ & $1.299\pm 0.045$ &
      $1.293\pm 0.045\pm 0.029$\\
      $R_1(1)$ & $1.643\pm 0.163\pm 0.112$ & $1.427\pm 0.050$ &
      $1.495\pm 0.050\pm 0.062$\\
      $R_2(1)$ & $0.842\pm 0.105\pm 0.038$ & $0.844\pm 0.034$ &
      $0.844\pm 0.034\pm 0.019$\\
      $R_{K3\pi/K\pi}$ & 2.153 \scriptsize{(fixed)} & & $2.153\pm
      0.011$\\
      $\mathcal{B}(B^0\to D^{*-}\ell^+\nu_\ell)$ (\%) & $4.47\pm
      0.04\pm 0.26$ & $4.43\pm 0.02$ & $4.42\pm 0.03\pm 0.25$\\
      $\mathcal{F}(1)|V_{cb}|$ (10$^{-3}$) & $35.6\pm 0.7\pm 1.3$ &
      $34.5\pm 0.2$ & $34.4\pm 0.2\pm 1.0$\\
      $\chi^2/$n.d.f. & 17.9/36 & & 138.8/155\\
      $P(\chi^2)$ & 99.5\% & & 82.0\%\\
      \hline\hline
    \end{tabular}
  \end{center}
  \caption{The results of the fits to the sub-samples, their average
  and the fit result on the total sample. The first error is
  statistical, the second is the estimated systematic uncertainty. The
  breakup of the systematic uncertainty is given in
  Table~\ref{tab:5}. All numbers are preliminary.} \label{tab:4}
\end{table}
\begin{table}
  \begin{center}
    \begin{tabular}{l|@{\extracolsep{.3cm}}ccccc}
      \hline\hline
      & $\mathcal{F}(1)|V_{cb}|$ & $\rho^2$ & $R_1(1)$ & $R_2(1)$ &
      $R_{K3\pi/K\pi}$\\
      \hline
      $\mathcal{F}(1)|V_{cb}|$ & 1.000 & 0.635 & $-0.285$ & $-0.220$ &
      0.011\\
      $\rho^2$ & & 1.000 & \phantom{$-$}0.388 & $-0.870$ &
      0.040\\
      $R_1(1)$ & & & \phantom{$-$}1.000 & $-0.511$ & 0.001\\
      $R_2(1)$ & & & & \phantom{$-$}1.000 & 0.002\\
      $R_{K3\pi/K\pi}$ & & & & & 1.000\\
      \hline\hline
    \end{tabular}
  \end{center}
  \caption{The statiscal correlation coefficients of the five
  parameters in the fit to the full sample.} \label{tab:5}
\end{table}

As explained earlier, the branching fraction for the decay $D^0\to
K^-\pi^+\pi^+\pi^-$ is floated in the fit to the full sample. The fit
result is compatible with the recent measurement by the CLEO-c
experiment~\cite{He:2005bs}. In the sub-sample fits, $R_{K3\pi/K\pi}$
is fixed to the value of the full sample fit.

\subsection{Systematic uncertainties}

To estimate the systematic uncertainties in the results quoted above,
we consider contributions from the following sources: uncertainties
in the background component normalizations, uncertainty in the MC tracking
efficiency, errors in $\mathcal{B}(D^{*+}\to D^0\pi^+)$ and
$\mathcal{B}(D^0\to K^-\pi^+)$~\cite{Yao:2006px}, and uncertainties in
the $B^0$~lifetime~\cite{Yao:2006px} and the total number of
$B^0$~mesons in the data sample.

To estimate the uncertainty related to a given background component,
we vary the normalization by $\pm 1$~standard deviation of the fit
described in Sect.~\ref{sec:3c}. The uncertainty in the
$D^{**}\ell\nu$~component is inflated by a factor of two to make the
amount of $D^{**}\ell\nu$ in each sub-sample consistent. The error in
the continuum normalization is taken to be 1.5\% as explained earlier.

For the tracking uncertainty, we calculate the track finding error
considering only the $D^0$~decay into $K^-\pi^+$, as the branching
fraction for $D^0\to K^-\pi^+\pi^-\pi^+$ is fitted from
data. (A possible mismodeling of the tracking efficiency for this mode
would be absorbed in $R_{K3\pi/K\pi}$.) We thus have four charged
tracks. Assuming 1\% uncertainty for each track, except for the
slow pion from $D^{*+}$ for which we use 2\%, and adding these
uncertainties linearly, we obtain a tracking uncertainty of
5\%. Given the size of this error, the uncertainty in the lepton
identification (1$-$2\%) can be neglected.

The breakup of the systematic error quoted in Table~\ref{tab:4} is
given in Table~\ref{tab:6}.
\begin{table}
  \begin{center}
    \begin{tabular}{l|@{\extracolsep{.3cm}}ccccc}
      \hline\hline
      & $\rho^2$ & $R_1(1)$ & $R_2(1)$ &
      $\mathcal{B}(D^*\ell\nu_\ell)$ & $\mathcal{F}(1)|V_{cb}|$\\
      \hline
      $D^{**}$ & 0.015 & 0.038 & 0.011 & 0.051 & 0.25\\
      uncorrelated & 0.009 & 0.028 & 0.002 & 0.003 & 0.04\\
      correlated & 0.003 & 0.003 & 0.007& 0.028 & 0.14\\
      fake $\ell$ & 0.020 & 0.037 & 0.009 & 0.002 & 0.04\\
      fake $D^{*}$ & 0.012 & 0.011 & 0.009 & 0.034 & 0.33\\
      continuum & 0.003 & 0.008 & 0.000 & 0.001 & 0.02\\
      tracking & $-$ & $-$ & $-$ & 0.221 & 0.86\\
      $\mathcal{B}(D^0\to K^-\pi^+)$~\cite{Yao:2006px} & $-$ & $-$ &
      $-$ & 0.081 & 0.31\\
      $\mathcal{B}(D^{*+}\to D^0\pi^+)$~\cite{Yao:2006px} & $-$ & $-$
      & $-$ & 0.033 & 0.13\\
      $\tau(B^0)$~\cite{Yao:2006px} & $-$ & $-$ & $-$ & 0.026 & 0.10\\
      $N(B\bar B)$ & $-$ & $-$ & $-$ & 0.036 & 0.14\\
      $f_{+-}/f_{0\bar 0}$~\cite{Yao:2006px} & 0.003 & 0.011 & 0.005 &
      0.001 & 0.04\\
      \hline
      total & 0.029 & 0.062 & 0.019 & 0.251 & 1.04\\
      \hline\hline
    \end{tabular}
  \end{center}
  \caption{The breakup of the systematic uncertainty in the result
  of the fit to the full sample.} \label{tab:6}
\end{table}

\section{Summary and discussion}

We have reconstructed about 69,000 $B^0\to
D^{*-}\ell^+\nu_\ell$~decays using a 140~fb$^{-1}$ data sample
recorded at the $\Upsilon(4S)$~resonance with the Belle detector at
the KEKB accelerator. A fit to four kinematic variables fully
characterizing this decays yields to measurements of the form factor
normalization $\mathcal{F}(1)|V_{cb}|$ and of the parameters $\rho^2$,
$R_1(1)$ and $R_2(1)$ that enter the HQET~form factor parameterization
of this decay. We obtain: $\mathcal{F}(1)|V_{cb}|=34.4\pm 0.2\pm 1.0$,
$\rho^2=1.293\pm 0.045\pm 0.029$, $R_1(1)=1.495\pm 0.050\pm 0.062$,
$R_2(1)=0.844\pm 0.034\pm 0.019$, and  $\mathcal{B}(B^0\to
D^{*-}\ell^+\nu_\ell)=(4.42\pm 0.03\pm 0.25)\%$. For all these
numbers, the first error is the statistical and the second is the
systematic uncertainty. These results are compatible with the recent
BaBar measurements of these quantities~\cite{Aubert:2007rs}. All
numbers are preliminary.

\section*{Acknowledgments}

We thank the KEKB group for the excellent operation of the
accelerator, the KEK cryogenics group for the efficient
operation of the solenoid, and the KEK computer group and
the National Institute of Informatics for valuable computing
and SINET3 network support. We acknowledge support from
the Ministry of Education, Culture, Sports, Science, and
Technology of Japan and the Japan Society for the Promotion
of Science; the Australian Research Council and the
Australian Department of Education, Science and Training;
the National Natural Science Foundation of China under
contract No.~10575109 and 10775142; the Department of
Science and Technology of India; 
the BK21 program of the Ministry of Education of Korea, 
the CHEP SRC program and Basic Research program 
(grant No.~R01-2005-000-10089-0) of the Korea Science and
Engineering Foundation, and the Pure Basic Research Group 
program of the Korea Research Foundation; 
the Polish State Committee for Scientific Research; 
the Ministry of Education and Science of the Russian
Federation and the Russian Federal Agency for Atomic Energy;
the Slovenian Research Agency;  the Swiss
National Science Foundation; the National Science Council
and the Ministry of Education of Taiwan; and the U.S.\
Department of Energy.


\begin{thebibliography}{99}

\bibitem{Kobayashi:1973fv}
  N.~Cabibbo,
  Phys.\ Rev.\ Lett.\  {\bf 10}, 531 (1963);

  M.~Kobayashi and T.~Maskawa,
  Prog.\ Theor.\ Phys.\  {\bf 49}, 652 (1973).

\bibitem{Briere:2002ew}
  R.~A.~Briere {\it et al.}  [CLEO Collaboration],
  Phys.\ Rev.\ Lett.\  {\bf 89}, 081803 (2002)
  [arXiv:hep-ex/0203032].

\bibitem{Abe:2001cs}
  K.~Abe {\it et al.}  [BELLE Collaboration],
  Phys.\ Lett.\  B {\bf 526}, 247 (2002)
  [arXiv:hep-ex/0111060].

\bibitem{Aubert:2007rs}
  B.~Aubert {\it et al.}  [BABAR Collaboration],
  Phys.\ Rev.\  D {\bf 77}, 032002 (2008)
  [arXiv:0705.4008 [hep-ex]].

\bibitem{ref:0} Throughout this note charge conjugation is implied.

\bibitem{Yao:2006px}
  W.~M.~Yao {\it et al.}  [Particle Data Group],
  J.\ Phys.\ G {\bf 33}, 1 (2006).

\bibitem{Neubert:1993mb}
  M.~Neubert,
  Phys.\ Rept.\  {\bf 245}, 259 (1994)
  [arXiv:hep-ph/9306320].

\bibitem{Isgur:1989vq}
  N.~Isgur and M.~B.~Wise,
  Phys.\ Lett.\  B {\bf 232}, 113 (1989).

\bibitem{Isgur:1989ed}
  N.~Isgur and M.~B.~Wise,
  Phys.\ Lett.\  B {\bf 237}, 527 (1990).

\bibitem{Richman:1995wm}
  J.~D.~Richman and P.~R.~Burchat,
  Rev.\ Mod.\ Phys.\  {\bf 67}, 893 (1995)
  [arXiv:hep-ph/9508250].

\bibitem{Hashimoto:2001nb}
  S.~Hashimoto, A.~S.~Kronfeld, P.~B.~Mackenzie, S.~M.~Ryan and J.~N.~Simone,
  Phys.\ Rev.\  D {\bf 66}, 014503 (2002)
  [arXiv:hep-ph/0110253].

\bibitem{Uraltsev:2000qw}
  N.~Uraltsev,
  arXiv:hep-ph/0010328.

\bibitem{Laiho:2007pn}
  J.~Laiho  [Fermilab Lattice and MILC Collaborations],
  PoS {\bf LATTICE2007}, 358 (2007)
  [arXiv:0710.1111 [hep-lat]].

\bibitem{Caprini:1997mu}
  I.~Caprini, L.~Lellouch and M.~Neubert,
  Nucl.\ Phys.\  B {\bf 530}, 153 (1998)
  [arXiv:hep-ph/9712417].

\bibitem{unknown:2000cg}
  A.~Abashian {\it et al.}  [Belle Collaboration],
  %
  Nucl.\ Instrum.\ Meth.\ A {\bf 479}, 117 (2002).

\bibitem{Kurokawa:2001nw}
  S.~Kurokawa,
  %
  Nucl.\ Instrum.\ Meth.\ A {\bf 499}, 1 (2003), and other papers
  included in this volume.

\bibitem{Lange:2001uf}
  D.~J.~Lange,
  %
  Nucl.\ Instrum.\ Meth.\ A {\bf 462}, 152 (2001).

\bibitem{Brun:1987ma}
  R.~Brun, F.~Bruyant, M.~Maire, A.~C.~McPherson and P.~Zanarini,
  %
  CERN-DD/EE/84-1.

\bibitem{Barberio:1993qi}
  E.~Barberio and Z.~Was,
  Comput.\ Phys.\ Commun.\  {\bf 79}, 291 (1994).

\bibitem{Abe:2001hj}
  K.~Abe {\it et al.}  [Belle Collaboration],
  %
  Phys.\ Rev.\ D {\bf 64}, 072001 (2001)
  [hep-ex/0103041].

\bibitem{Fox:1978vu}
  G.~C.~Fox and S.~Wolfram,
  Phys.\ Rev.\ Lett.\  {\bf 41}, 1581 (1978).

\bibitem{Hanagaki:2001fz}
  K.~Hanagaki, H.~Kakuno, H.~Ikeda, T.~Iijima and T.~Tsukamoto,
  Nucl.\ Instrum.\ Meth.\ A {\bf 485}, 490 (2002)
  [hep-ex/0108044].

\bibitem{Abashian:2002bd}
  A.~Abashian {\it et al.},
  Nucl.\ Instrum.\ Meth.\ A {\bf 491}, 69 (2002).

\bibitem{ref:2} Quantities evaluated in the c.m.\ frame are denoted by
  an asterisk.

\bibitem{Barlow:1993dm}
  R.~J.~Barlow and C.~Beeston,
  Comput.\ Phys.\ Commun.\  {\bf 77}, 219 (1993).

\bibitem{Brun:1997pa}
  R.~Brun and F.~Rademakers,
  Nucl.\ Instrum.\ Meth.\  A {\bf 389} (1997) 81.

\bibitem{James:1975dr}
  F.~James and M.~Roos,
  Comput.\ Phys.\ Commun.\  {\bf 10}, 343 (1975).

\bibitem{He:2005bs}
  Q.~He {\it et al.}  [CLEO Collaboration],
  Phys.\ Rev.\ Lett.\  {\bf 95}, 121801 (2005)
  [Erratum-ibid.\  {\bf 96}, 199903 (2006)]
  [arXiv:hep-ex/0504003].

\end{thebibliography}
\end{document}